\documentclass[fleqn,usenatbib]{mnras}

\usepackage{newtxtext,newtxmath}

\usepackage[T1]{fontenc}

\DeclareRobustCommand{\VAN}[3]{#2}
\let\VANthebibliography\thebibliography
\def\thebibliography{\DeclareRobustCommand{\VAN}[3]{##3}\VANthebibliography}

\usepackage{graphicx}	
\usepackage{amsmath}	
\usepackage{subcaption}
\usepackage{xcolor}

\setcitestyle{citesep={,}}
\newcommand{\degree}{$^{\circ}$}

\usepackage{mycommands}
\usepackage{tablefootnote}

\title[Clustering of AGN/SFGs in LoTSS Deep Fields]{The Clustering of Active Galactic Nuclei and Star Forming Galaxies in the LoTSS Deep Fields}

\author[C. L. Hale et al.]{C. L. Hale,$^{1,2}$\thanks{E-mail: catherine.hale@physics.ox.ac.uk}
P. N. Best,$^{2}$
K.J.\ Duncan,$^{2}$
R. Kondapally,$^{2, 3, 4}$
M. J. Jarvis,$^{1, 5}$
M. Magliocchetti,$^{6}$ \newauthor
H. J. A. R\"ottgering,$^{7}$
D. J.   Schwarz,$^{8}$
D J. B. Smith,$^{9}$
and J. Zheng$^{8}$
\\
$^{1}$Astrophysics, Denys Wilkinson Building, Department of Physics, University of Oxford, Keble Road, OX1 3RH, UK\\
$^{2}$Institute for Astronomy, School of Physics and Astronomy, University of Edinburgh, Royal Observatory, Blackford Hill, Edinburgh, EH9 3HJ, UK \\
$^{3}$ Centre for Extragalactic Astronomy,Department of Physics, Durham University, Durham DH1 3LE, UK \\
$^{4}$ Centre for Extragalactic Astronomy,Department of Physics, Durham University, Durham DH1 3LE, UK \\
$^{5}$ Department of Physics and Astronomy, University of the Western Cape, Robert Sobukwe Road, 7535 Bellville, Cape Town, South Africa. \\
$^{6}$ INAF-IAPS, Via del Fosso del Cavaliere 100, 00133 Rome, Italy \\
$^{7}$ Leiden Observatory, Leiden University, PO Box 9513, 2300 RA Leiden, The Netherlands \\
$^{8}$ Fakult\"at f\"ur Physik, Universit\"at Bielefeld, Postfach 100131, 33501 Bielefeld, Germany \\
$^{9}$ Centre for Astrophysics Research, University of Hertfordshire, College Lane, Hatfield AL10 9AB, UK \\
}

\date{Accepted XXX. Received YYY; in original form ZZZ}

\pubyear{2024}

\begin{document}
\label{firstpage}
\pagerange{\pageref{firstpage}--\pageref{lastpage}}
\maketitle

\begin{abstract}
Using deep observations across three of the LOFAR Two-metre Sky Survey Deep Fields, this work {measures} the angular clustering of star forming galaxies (SFGs) and low-excitation radio galaxies (LERGs) to $z$$\lesssim$1.5 {{for} faint {sources}, $S_{\textrm{144 MHz}}$$\geq$200 $\muup$Jy}. We {measure} the angular auto-correlation {of LOFAR} sources in redshift bins and their cross-correlation with multi-wavelength sources {to} measure the evolving galaxy bias for SFGs and LERGs. {{Our} work shows the bias {of the radio-selected SFGs increases} from {$b$=$\bccSFGOne^{+\bccperrSFGOne}_{-\bccnerrSFGOne}$ at $z$$\sim$0.2 to $b$=$\bccSFGFive^{+\bccperrSFGFive}_{-\bccnerrSFGFive}$} at $z$$\sim$1.2; {faster than the assumed $b(z)$$\propto$$1/D(z)$ models}} adopted in {previous} LOFAR cosmology studies (at sensitivities where AGN dominate), but in broad agreement with previous work. {We further study the luminosity dependence of bias for SFGs and find little evidence for any luminosity dependence at fixed redshift, although uncertainties remain large for the sample sizes available}. The LERG population instead {shows a weaker {redshift} evolution} with {$b$=$\bccLERGOne^{+\bccperrLERGOne}_{-\bccnerrLERGOne}$ at $z$$\sim$0.7 to $b$=$\bccLERGTwo^{+\bccperrLERGTwo}_{-\bccnerrLERGTwo}$} at $z$$\sim$1.2, {though it is also consistent with the assumed bias evolution model {($b(z)$$\propto$$1/D(z)$)} {within the measured uncertainties}}. For those LERGs which reside in quiescent galaxies (QLERGs), there is weak evidence {that they are more biased than} the general LERG population {and evolve from $b$=$\bccQLERGOne^{+\bccperrQLERGOne}_{-\bccnerrQLERGOne}$ at $z$$\sim$0.7 to $b$=$\bccQLERGTwo^{+\bccperrQLERGTwo}_{-\bccnerrQLERGTwo}$ at $z$$\sim$1.2}. This {suggests} the halo environment of radio sources {may be related} to their properties. {These measurements} can help {constrain} models for the bias evolution of these source populations, and can help {inform} multi-tracer {analyses}. 
\end{abstract}

\begin{keywords}
cosmology: large-scale structure of Universe, observations -- radio continuum: galaxies
\end{keywords}



\section{Introduction}
\label{sec:intro}

{Large-area} spectroscopic surveys have been instrumental in allowing us to observe how galaxies are distributed and to build up knowledge of the cosmic web. These {surveys} demonstrate that galaxies are not uniformly distributed and that there is large-scale structure within the Universe. Surveys such as the 2dF Galaxy Redshift Survey \citep[2dFGRS;][]{2dfgrs}, 6dF Galaxy Survey \citep[6dFGS;][]{6dFGRS}, Sloan Digital Sky Survey \citep[SDSS;][]{sdss} and Galaxy And Mass Assembly \citep[GAMA;][]{gama} survey, have all been crucial in making detailed maps of the distribution of galaxies in the Universe, though many of these were limited to more local structures $z<1$. These observations {show} clusters filled with galaxies, filaments connecting the clusters, and regions with a clear deficit of galaxies, known as voids. 

By studying how different galaxies are distributed within this cosmic web, we are able to gain greater understanding of the impact of the underlying environments on galaxies and their properties. This distribution of galaxies in the large-scale structure can be {studied through} the spatial two-point correlation function, $\xi(r)$, \citep[see e.g.][]{Peebles1980}. $\xi(r)$ quantifies the excess probability to find {galaxy pairs} at a {comoving} spatial scale ($r$), compared to if {galaxies are} randomly distributed in the Universe. More formally, $\xi(r)$ is defined by:

\begin{equation}
dP(r) = \overline{n} \left[ 1 + \xi(r) \right] d^3r,
\label{eq:xi}
\end{equation}

\noindent where $\overline{n}$ is the mean density of sources and $dP$ the probability to observe galaxies in a volume, $d^3r$, at a given spatial separation, $r$. The spatial clustering of the aforementioned spectroscopic surveys has been studied in great detail and {allows {the properties} of galaxies to be} related to their underlying dark matter environments \citep[in numerous works including][]{Madgwick2003, Zehavi2011, Guo2015}.

However, for the majority of surveys both at radio wavelengths and across the electromagnetic spectrum, we are unable to obtain spectroscopic completeness for large area, deep surveys of galaxies. This is especially true for the {radio surveys} which are being {carried out} with state-of-the-art radio facilities such as LOFAR \citep{vanHaarlem2013}, ASKAP \citep{ASKAP} and MeerKAT \citep{Jonas2009, Jonas2016}. Continuum surveys from such facilities instead image galaxies at specific {frequencies,} {and} cannot directly provide redshift information. Instead radio surveys rely on counterpart sources from other {wavebands across} the electromagnetic spectrum to determine redshifts. Where spectroscopic redshifts are unavailable, photometric redshifts are relied upon. These photometric redshifts {combine the available multi-wavelength} data and use template fitting \citep[e.g.][]{LePHARE, Eazy} or machine learning methods \citep[see e.g.][]{Almosallam2016a, Almosallam2016b, METAPHOR} to assign redshifts ($z$). Such redshifts can have broad probability density functions (PDFs), {{arising from} uncertainties in modelling photometric redshifts using the data available, and the distributions can often have multiple peaks}. The spatial distribution of samples where photometric redshifts dominate are, therefore, much more uncertain. However, we are still able to gain {an} understanding {of} the distribution of galaxies using their projected clustering by measuring the angular two-point correlation function {(TPCF)}, $\omega(\theta)$, {defined by:}
\begin{equation}
dP(\theta) = \sigma \left[ 1 + \omega(\theta) \right] d\Omega.
\end{equation}

\noindent This is similar to Equation \ref{eq:xi}, where now $dP$ is the probability to observe galaxies within angular separations ($\theta$){,} $\sigma$ is the average surface density of sources, and $d\Omega$ is the solid angle element being considered. 

In practice, $\omega(\theta)$ is calculated from galaxy surveys using estimators \citep[such as from][]{Hamilton1993,LandySzalay} through comparing counts of galaxies within angular separations compared to randomly distributed galaxies. This does not rely upon any redshift information. However, using the {overall} redshift distribution of the sources, and {assuming a model for $\xi(r)$, the spatial clustering can be inferred \citep[see e.g. Limber inversion and its use in a number of radio studies; ][]{Limber1953, Limber1954, Peebles1980, Overzier2003, Lindsay2014, Hale2018}.} Knowledge of the clustering and the redshift distribution of sources can be further used to relate {their} clustering to that of the spatial clustering of the underlying matter, $\xi_{m}$. This allows quantification for a parameter known as {bias, $b$, \citep[see e.g. discussions in][]{Peebles1980, Peacock2000, Desjacques2018}, defined by:}
\begin{equation}
\xi(z, r) = b^2(z, r) \xi_{m}(z, r).
\end{equation}
Through tracing how bias evolves for a population of {sources, the} relationship between galaxies, their properties and the underlying matter environment {can} be studied to better quantify the evolving galaxy-halo connection.

The angular two-point correlation function has been relied upon for a number of studies into the clustering of radio sources. These cover both wide area surveys \citep[e.g.][]{Blake2002, Overzier2003, sumssclustering} and smaller regions over which there is deep ancillary data \citep[e.g.][]{Lindsay2014, Magliocchetti2017, Hale2018, Chakraborty2020}. {Recent studies with LOFAR have also been used to probe the clustering of radio detected sources and study the relationship of such galaxies to their dark matter environment; however they have typically focussed on bright populations \citep[$S_{\textrm{144 MHz}}$$\gtrsim$2 mJy;][]{Siewert2020, Alonso2021, Hale2024, Nakoneczny2024, Bhardwaj2024}.} Whilst {radio clustering studies often rely on the} angular clustering due to the dominance of photometric redshifts, this will be improved upon with future spectroscopic surveys which specifically target {the host galaxies of} radio detected sources \citep[see e.g.][]{Smith2016, ORCHIDSS, Jin2023}.

With the radio observations of recent, deeper surveys with telescopes such as LOFAR \citep{Williams2016, Hale2019, Tasse2021, Sabater2021}, MeerKAT \citep[e.g.][]{Mauch2020,Heywood2021, Hale2024b}, ASKAP \citep{Norris2021, Gurkan2022}, u-GMRT \citep[e.g.][]{Mazumder2020, Ocran2020} and the VLA \citep[e.g.][]{Smolcic2017, vandervlugt2021}, we are in the regime where star forming galaxies (SFGs) contribute a significant fraction to, and can even dominate, the total source population {\citep[see e.g.][]{Smolcic2017b, Algera2020, Best2023}}. Using such surveys that combine area, sensitivity and have a wealth of ancillary data it is possible to identify host galaxies for these radio sources and classify these sources into different sub-classes \citep[e.g. active galactic nuclei, AGN, and SFGs, see e.g.][]{Smolcic2017b, Algera2020, Whittam2022, Best2023, Das2024}. This classification allows for in depth studies of the statistical properties of different source populations and their connection to their host properties, environments and redshifts. 

{Furthermore, AGN {can be} further categorised based on their properties. Historically, AGN have both been split based on morphological properties \citep{Fanaroff1974} as well as into radio `loud' and `quiet' populations which distinguish the significance {of} the radio emission from the jets \citep[see e.g.][]{Wilson1995}. For radio loud AGN (RLAGN), these are often further split based on their {accretion onto the central AGN}, which may occur in two fundamental modes based on their radiative efficiency \citep[see e.g.][which provide reviews on this topic]{Heckman2014, Hardcastle2020}. {Those radio} sources which accrete from radiatively efficient disks, {are} known as high excitation radio galaxies (HERGs) {and} are believed to be geometrically thin, optically thick accretion disks \citep[][]{Shakura1973}. Conversely, low excitation radio galaxies (LERGs) are believed to accrete from a radiatively inefficient disk, which {are thought to be fuelled by} advection dominated flows \citep{Narayan1994, Narayan1995}. {{However, recent studies such as those from \cite{Whittam2018, Whittam2022} have indicated a greater overlap in the {accretion} efficiency of these two populations.} } }

Previous clustering studies in the radio have shown that different source populations cluster differently, with AGN found to be, in general, more highly clustered than their star formation dominated counterparts \citep[see e.g.][]{Magliocchetti2017, Hale2018, Chakraborty2020,Mazumder2022}. Owing to this, it is important for radio clustering studies to study the evolution of different source populations independently. {It is also important to understand the clustering of different source populations so that their bias can be} applied to multi-tracer techniques {to help overcome cosmic variance at large scales} \citep[see e.g.][]{Raccanelli2012, Ferramacho2014, Gomes2020}. Recent work has also indicated that there may be a connection between the accretion mode of radio loud AGN and {their} clustering, through the study of high redshift analogues of high/low-excitation radio galaxies \citep[H/LERGs; see][]{Hale2018}. A recent summary of a number of radio-based clustering studies can be found in \cite{Magliocchetti2022}. {Moreover, other studies which probe the environments of H/LERGs {through other measurements} have also suggested differences in their local environments may be important \citep[see e.g.][]{Tasse2008, Gendre2013, Hardcastle2020}.}

In this work we make use of some of the deepest LOFAR observations to date and complementary multi-wavelength data to study the clustering of SFGs and LERGs in three of the LOFAR Two-metre Sky Survey (LoTSS) Deep Fields. {SFGs and LERGs represent the two most populous source types identified in the LoTSS deep fields \citep{Best2023}.} This paper is presented as follows. In Section \ref{sec:data} we present the data used in this work both from LOFAR and the associated multi-wavelength catalogues. In Section \ref{sec:tpcf} we outline the methods used to measure the angular clustering of {the} SFGs and LERGs and we present the results of such analysis in Section \ref{sec:omega}. We then discuss the galaxy bias and present its evolution in Section \ref{sec:bias}. Furthermore, owing to the larger radio samples available from the LOFAR Deep Fields, we further investigate the evolving bias as a function of luminosity for LOFAR detected SFGs. We present the conclusions of our analysis in Section \ref{sec:conclusions}. Unless otherwise stated, our work assumes a constant spectral index for radio sources, $\alpha=0.7$, where $S_{\nu} \propto \nu^{-\alpha}$ and we adopt the cosmology used in \cite{Kondapally2021} and \cite{Best2023}, namely: $H_0=70$ km s$^{-1}$ Mpc$^{-1}$, $\Omega_m=0.3$, $\Omega_{\Lambda}=1-\Omega_m$ and also adopt $n_s=0.965$ and $\sigma_8=0.8$.

\begin{table*}
    \begin{tabular}{l | c c c | c c c}
       \textbf{Description}  & \multicolumn{3}{| c |}{ \textbf{Radio Catalogue}} &  \multicolumn{3}{| c| }{ \textbf{Multi-Wavelength Catalogue}} \\ 
       & Bo\"otes & ELAIS-N1 & Lockman & Bo\"otes & ELAIS-N1 & Lockman \\ \hline \hline
       Original Source Catalogue & 36 767 &  {70 544} & 50 112 & {2 214 329} & {2 105 993} &  {3 041 793} \\[4pt] 
        In FLAG\_OVERLAP and FLAG\_CLEAN regions & 18 553 &  30 768 & 30 347 &  {1 911 265} & {1 446 319} & {1 837 134} \\[1pt] 
       \ \ \ \ \ \ \citep[+ (for radio) in cross-matched catalogue of][]{Kondapally2021} & & & & & & \\[4pt]  
      Band Used for magnitude and 5$\sigma$ Cut & - & - &  - & 4.5 $\muup$m & \textit{K} & 4.5 $\muup$m  \\[4pt] 
       Magnitude Cut Applied & - & - &  - & $\leq$21.33 & $\leq$21.78 & $\leq$21.18 \\[4pt] 
       Applying 5$\sigma$ Cut and Magnitude Cuts& -& - &  - & 317 022 & 282 871 & 302 530 \\[4pt] 
       With redshifts from \protect \cite{Duncan2021}  & 18 238 & 30 470 & 30 161 & 216 708 & 272 315 & 297 071 \\[1pt] 
       With source classification from \protect \cite{Best2023}  & 17 707 & 30 182 & 29 595 & - & - & - \\[1pt] 
       Additional spatial masking applied (additional star masks & 15 905 & 28 772 & 27 977 & 210 714 & 260 949 & 284 576 \\[1pt]   
        \ \ \ \ \ \  for all fields + masking of Table \ref{tab:mask_bootes} for Bo\"otes)& & & & & & \\[4pt]  
       Radio flux density ($\geq$200$\muup$Jy) \& SNR ($\geq$5$\sigma$) cuts & \textbf{14 925} & \textbf{17 289} & \textbf{22 797} & - & - & - \\[4pt] 
       Mass cut applied ($M_{*}\geq 10^{10.5} M_{\odot}$)& - & - & - & \textbf{68 257} & \textbf{59 636} & \textbf{87 525} \\[4pt] 
    \end{tabular}
    \caption{{Table outlining the number of sources (across all redshift {ranges) from} the initial catalogues and after applying {the} subsequent cuts which are used in this analysis. Each row is cumulative and includes the cuts applied to all previous rows. These numbers are indicated for the radio and multi-wavelength catalogues separately, for each of the three fields (Bo\"otes, ELAIS-N1 and Lockman Hole) and are not split by source type (i.e. SFG vs LERG). Those numbers indicated in bold text show the final numbers of either radio or multi-wavelength sources used across all redshifts. Those used for each redshift sample considered in this work can be found in \protect Table \ref{tab:bias_tab}. We note that in this table, the starting criteria ``Original Catalogue" for the radio populations refers to the source catalogue over the $\sim$25 deg$^2$ of each field within the primary beam cut, as described in \protect \cite{Tasse2021, Sabater2021}. For the multi-wavelength data the full catalogue relates to the `Science Ready' catalogues released with \protect \cite{Duncan2021}. } }
\label{tab:Nsources}
\end{table*}

\section{Data}
\label{sec:data}
The data used in this work come from LOFAR observations across the LoTSS Deep Fields and their {associated value added catalogues}. We summarize the data here but comprehensive details can be found in \cite{Sabater2021, Tasse2021} for the radio continuum {images and catalogues} and \cite{Kondapally2021, Duncan2021, Best2023} for the host galaxy identification, {redshift estimation and {source classification}}, respectively.

\subsection{Radio Data: LoTSS Deep Fields}
\label{sec:data_lotssdeep}
The LoTSS Deep Fields consist of four well-studied multi-wavelength fields: Bo\"otes, Lockman Hole, the European Large-Area ISO Survey Northern Field 1 (ELAIS-N1) and the North Ecliptic Pole (NEP) field. These fields are all located in the northern sky, at optimal locations for LOFAR (which is not a {physically} steerable telescope) {to observe}. Using the high band antenna (HBA) of LOFAR, {the first observations at} 144 MHz of three of these deep fields were {published} in a combination of papers for ELAIS-N1 \citep{Sabater2021} and the Bo\"otes and Lockman Hole \citep{Tasse2021} fields. \cite{Sabater2021} and \cite{Tasse2021} presented images and catalogues for {observations totalling} 164, 80 and 112 hours on target for the ELAIS-N1, Bo\"otes and Lockman Hole fields respectively, covering approximately 25 deg$^2$ in each field\footnote{{Where {the $\sim$25 deg$^2$ corresponds to the area of} the images released in \protect \cite{Tasse2021} and \protect \cite{Sabater2021} {which are} truncated at the 30\% power point of the primary beam. \texttt{PyBDSF} was run for each image over this {full $\sim$}25 deg$^2$.}}. 

Processing of the data used a combination of flagging and averaging of the raw dataset, then calibrating the data. This calibration consisted of both direction-independent and direction-dependent calibration, making use of the packages \texttt{kMS} \citep{Tasse2014, Smirnov2015, Tasse2023kMS} and \texttt{DDFacet} \citep{Tasse2018, Tasse2023DDF}. Direction-dependent calibration is crucial for observations at such low frequencies to account for the effects of the ionosphere, which can cause the apparent movements of sources across the sky, but also is necessary to account for primary beam effects {over} long duration observations. These direction-dependent corrections allow for {images with an angular resolution of} 6\arcsec {to be produced}, compared to {25\arcsec} {with direction-independent calibration alone} \citep[see][]{Shimwell2017}.

Source catalogues were extracted in each of the fields using the Python Blob Detection Source Finder \citep[\texttt{PyBDSF}; ][]{Mohan2015}, using a 5$\sigma$ peak signal-to-noise thresholding {criterion}. Owing to its longer observations, ELAIS-N1 is the deepest field with an average rms of $\sim$30 $\muup$Jy beam$^{-1}$ across the image. This compares to $\sim$60 $\muup$Jy beam$^{-1}$ in Bo\"otes and $\sim$40 $\muup$Jy beam$^{-1}$ in the Lockman Hole field. This results in the detection of a total of $\sim$157,000 sources across the $\sim$25 deg$^2$ of radio area in each of the three fields, with $\sim$70 000, $\sim$37 000, $\sim$50 000 sources in the ELAIS-N1, Bo\"otes and Lockman Hole fields respectively. {{In this paper we will adopt a subset of these catalogues for the analysis, we describe such cuts to the data in the following sections.}}

\subsection{Multi-Wavelength Data}
\label{sec:optIRdata}

Alongside the radio data, we make use of the multi-wavelength catalogues of sources detected in the three fields. These were not only used to provide counterparts to the radio sources (see Section \ref{sec:data_optIR_counterparts}) but are also {used} here to measure the angular cross-correlation {of these multi-wavelength galaxies} with the radio sources. These catalogues combine data from the UV to the far-IR and are described in detail in \cite{Kondapally2021}; their overlapping regions cover a reduced area compared to the radio data alone \citep[see Fig. 1 of][{where we make use of their shaded regions for this work}]{Kondapally2021}. 

For the Bo\"otes field, {the} multi-wavelength catalogue originates from 4.5 $\muup$m and I-band point spread function (PSF) matched catalogues from \cite{Brown2007, Brown2008} which combine data from the NOAO Deep Wide Field Survey \citep[NDWFS;][]{Jannuzi1999} as well as optical imaging from \cite{Bian2013} and near-IR data from \cite{Gonzalez2010}. For ELAIS-N1 and the Lockman Hole field, \cite{Kondapally2021} created their own combined matched-aperture catalogues. This includes data from the UV to IR: the Galaxy Evolution Explorer (GALEX) space telescope \citep{Martin2005, Morrissey2007}; Hyper-Suprime-Cam Subaru Strategic Program (HSC-SSP) survey \citep{Aihara2018}; the Canada France Hawaii Telescope (CFHT) MegaCam instrument \citep{Hildebrandt2016}; Panoramic Survey Telescope and Rapid Response System \citep[Pan-STARRS-1;][]{Chambers2016}; the Herschel Space Observatory \citep{Griffin2010, Poglitsch2010} and from the Spitzer Space Telescope \citep[from][]{Lonsdale2003, Mauduit2012}. \cite{Kondapally2021} {generated 0.2\arcsec \ pixel scale images} and detected sources using \textsc{SExtractor} \citep{Bertin1996} {to create the multi-wavelength catalogues which we use in this work}. Aperture corrections are additionally applied to account for varying point spread function (PSF) sizes between the images.

These combined {multi-wavelength} catalogues contain over 2 million {sources in each of the three fields used in this work}: $\sim$2.1 million in ELAIS-N1, $\sim$3.0 million in Lockman Hole and $\sim$2.2 million in Bo\"otes. This is reduced in numbers when only the areas which have overlap between all the best multi-wavelength surveys are considered and masking is applied \citep[see Fig 1 of][]{Kondapally2021}. This overlapping area covers $\sim$26 deg$^2$ {in total} across the three fields and reduces the number of multi-wavelength sources to {$\sim$1.4 million} sources in ELAIS-N1 (6.74 deg$^2$), {$\sim$1.9 million} sources in the Lockman Hole field (10.28 deg$^2$) and {$\sim$1.8 million} sources in Bo\"otes (8.63 deg$^2$). {For full details {of the sources used after cuts to the catalogues are applied} see Table \ref{tab:Nsources}.}  {The {sources in the areas adopted are identified using} a combination of the `FLAG\_OVERLAP' (==1 for Bo\"otes, ==3 for Lockman Hole and ==7 for ELAIS-N1) and `FLAG\_CLEAN' (!=3) identifiers in the source catalogue. {This restricts} the data to the best multi-wavelength regions, avoiding objects such as stars which may be impacting the multi-wavelength {photometry}. {Further details of the flag can be found in \cite{Kondapally2021}\footnote{{and in the read me files available at \url{https://lofar-surveys.org/index.html}}}, where FLAG\_OVERLAP is indicative of the availability of multi-wavelength coverage in different bands and FLAG\_CLEAN relates to the masking around bright stars.} In Bo\"otes an additional flag to mask ultra deep regions is also applied: `FLAG\_DEEP'(==1).}  

{Finally, we} also apply a stellar mass cut of M$_{\star}\geq$10$^{10.5}$ M$_{\odot}$ to the {multi-wavelength} data which, as can be seen in Fig. 11 of \cite{Duncan2021}, {is predominately larger than the 90\% {magnitude completeness limits already applied in this work}}. {Applying a {constant, high} mass cut is {generally} more restrictive than using the magnitude cuts of \cite{Duncan2021} {alone} to impose {completeness. {As discussed in \cite{Duncan2021}, the stellar masses {in their catalogue} are believed to be robust {up} to a redshift of $z=1.5$ and so we restrict ourselves to such a redshift range over which to probe the clustering}.  {The result of such a high mass cut is a robust sample of massive galaxies for cross-correlating to the radio data}}}. {{The high stellar mass cut also allows restricts the samples to} the most massive galaxies, which {is} beneficial when considering the {angular} cross-correlation, due to {the} larger bias \citep[see e.g.][]{Hatfield2016}. {Finally, it} also ensures that a similar reference sample of galaxies is {considered} across the redshift samples used in this work} {as well as between the three fields, to ensure we cross-correlate to a similar population}. {Such cuts reduced the number of {multi-wavelength} sources across the three fields. For more details on the effect of source numbers on the cuts applied see Table \ref{tab:Nsources}. }

\subsection{Radio {Data:} Host Identification and Redshifts}
\label{sec:data_optIR_counterparts}

In order to obtain redshift information and source classifications for the radio detected galaxies, a catalogue of multi-wavelength counterparts with redshifts are essential. The cross-matching process for the LoTSS Deep Fields is described in \cite{Kondapally2021}, where a combination of likelihood ratio {(LR)} analysis \citep[see e.g.][]{Sutherland1992, McAlpine2012, Williams2019, Whittam2024} as well as visual classification was used to identify the host for the radio sources as in \cite{Williams2019}\footnote{For clarity, prior to source association a radio source refers to the source as defined by the source finder, \texttt{PyBDSF}. After source associations and {classifications a source {refers to the object within the catalogue of \cite{Kondapally2021}. This is} assumed to be from an individual galaxy, which may include multiple of the original \texttt{PyBDSF} radio sources}.}. Due to the availability of multi-wavelength data, a restricted region of the three fields was used for the host identification process, as discussed in Section \ref{sec:optIRdata} {and presented in Table \ref{tab:Nsources}. Over} these smaller, multi-wavelength {areas which are} closer to the primary beam centre, the sensitivity improves, now measuring a typical rms of $\sim$20 $\muup$Jy beam$^{-1}$ in ELAIS-N1, $\sim$30 $\muup$Jy beam$^{-1}$ in Lockman Hole and $\sim$40 $\muup$Jy beam$^{-1}$ in Bo\"otes.

\cite{Kondapally2021} used a number of decision trees in order to identify which sources had a reliable identification of a host galaxy from the likelihood ratio analysis, and which sources instead needed visual identification {to obtain a host galaxy match}. Sources with compact sizes or secure radio positions were determined to be most suitable for LR cross-matching; sources with a large size or in a crowded region of the field were instead sent to visual analysis. {For visual} identification, \cite{Kondapally2021} used the Zooniverse\footnote{\url{https://www.zooniverse.org}} platform \citep[see e.g.][]{Zooniverse, Banfield2015, Williams2019} where LOFAR {surveys} team members used the interface to help visually cross-match sources to a host galaxy. Each source required {at least} five {independent classifications and a host galaxy was assigned if} at least 60 per cent of users who analysed a source agreed on a match. Sources without a clear match or that were flagged as requiring further detailed inspection were sent to experts for assessment. Approximately 97\% of the LOFAR detected sources within the multi-wavelength region have host galaxies identified.

{Alongside the work of \cite{Kondapally2021},} \cite{Duncan2021} used the wealth of multi-wavelength data across the three LoTSS Deep Fields to obtain redshift {estimates for the host galaxies. These redshifts {are} a combination of photometric redshifts {and} spectroscopic redshifts, where available}. Photometric redshifts were generated through a hybrid method which combines redshifts from spectral energy distribution (SED) fitting techniques and machine learning methods \citep[GPz;][]{Almosallam2016a, Almosallam2016b}.  
This method produced redshifts for as many sources as possible in the full multi-wavelength catalogue (described in Section \ref{sec:optIRdata}), which in turn can provide redshifts for a number of LOFAR detected sources. 
{In} total, 21 per cent of cross-matched sources in the Bo\"otes field have spectroscopic redshifts, reducing to 5 per cent in the other two fields at the time of release. Further details of these catalogues can be found in \cite{Duncan2021}. We note that {additional} spectra have subsequently been obtained with the Dark Energy Spectroscopic Instrument \citep[DESI;][]{Adame2023, DESI2025}, {however these were not available when host identification was conducted and when the redshifts were used to help in the classification of sources in \cite{Best2023}. We therefore rely on the redshift {information from} \cite{Duncan2021}}.   

\subsubsection{Additional Spatial Masking}
\label{sec:mask}

{{We} apply additional spatial cuts {to remove some remaining non-uniformity}. Firstly, we apply additional spatial masking in the Bootes field, {detailed} in Table \ref{tab:mask_bootes}}. {This {{avoids areas} in the Bo\"otes field that appeared deeper than the surrounding image and this was depth was not removed by use of FLAG\_DEEP in the catalogue.} {Secondly, {{we expand the stellar} masks of \cite{Kondapally2021} {to provide more}} conservative masking around the brightest stars in the Gaia catalogue \citep{Gaia, Gaia2}.} This is to ensure uniformity in the optical catalogues close to bright stars. We {create} a mask {around} sources with magnitudes in the G band $\leq10$ of 2\arcmin \ and 4\arcmin \ for those source with G band magnitudes $\leq7.5$. This {masks} 42, 61 and 64 stars across the Bo\"otes, ELAIS-N1 and Lockman Hole fields respectively {and} removes an additional 1-2\% {of sources in the} original cross-matched catalogue of \cite{Kondapally2021} compared to {the flagged regions discussed earlier of \cite{Kondapally2021}}. {Such spatial cuts {were} applied to both the radio and {multi-wavelength} data as well as the random catalogues.}}

\begin{table}
\centering
    \begin{tabular}{ c| c c}
    Region & RA Range ($^{\circ}$) & Dec Range ($^{\circ}$) \\ \hline
        1 & 218.90 - 219.00 & 33.45 - 33.70 \\
        2 & 216.10 - 219.00 & 33.45 - 33.53 \\
        3 & 218.20 - 218.30 & 32.85 - 33.70 \\
        4 & 216.10 - 218.30 & 32.85 - 32.96 \\
        5 & 217.50 - 217.60 & 32.32 - 33.68 \\
        6 & 216.78 - 216.90 & 32.32 - 33.68 \\
    \end{tabular}
    \caption{Additional regions within the Bo\"otes field which {are masked within the field}, {see Section \ref{sec:randoms_optIR}}.}
    \label{tab:mask_bootes}
\end{table}

\subsection{Source Classifications}
\label{sec:data_class}
For the cross-matched sources, classifications were determined using the abundance of multi-wavelength data and were released in \cite{Best2023}. In their work, \cite{Best2023} used a combination of {SED} fitting codes to assign classifications for the sources. This included the SED fitting codes \textsc{AGNFitter} \citep{AGNfitter}, \textsc{Bagpipes} \citep{Bagpipes}, \textsc{CIGALE} \citep{Cigale} and \textsc{MAGPHYS} \citep{Magphys1, Magphys2} to provide source properties for the host {galaxies.} For sources with an identified host, these were classified as either a star forming galaxy (SFG) or an active galactic nucleus (AGN). For those classified as an AGN, these were sub-classified as either Radio loud (RL) or Radio quiet (RQ) and for those RLAGN, these were classed as either {HERGs or LERGs}. \cite{Best2023} present a consensus classification for {the} majority of the LOFAR cross-matched sources, {whilst $\sim$1500 sources per field remain unclassified \citep[see Table 2 of][]{Best2023}}. This is a small fraction of the {total sources within the multi-wavelength region, $\sim$5\%, and this number includes those sources without an assigned host galaxy and redshift}. Further details of the classification methods used are provided in \cite{Best2023}. {For the classified population, approximately 68\% of sources within the multi-wavelength region were identified as SFGs, with LERGs being the next {biggest fraction} of sources at $\sim$16\%.}  \\ 

\noindent Owing to SFGs and LERGs being the two {largest} populations in the LoTSS Deep Fields, we investigate the clustering of these two populations in this work. We also {study the clustering for a subset of the LERG population, namely} Quiescent LERG  (or QLERGs). These are discussed in \cite{Kondapally2022} and are useful to this work as they provide a more representative comparison to the LERG population {used in the clustering {work} of \cite{Hale2018}, {who measure the clustering from a sample of quiescent LERGs from the catalogues from \cite{Smolcic2017b}}. We use the same criterion as in \cite{Kondapally2022} to classify sources as QLERGs, namely making cuts based on the specific star formation rate of the sources}. {We note, though, that alternative classifications for the ELAIS-N1 field were subsequently presented in \cite{Das2024} using the SED fitting code, \texttt{PROSPECTOR} \citep{Leja2017, Johnson2021}. Comparison of the ELAIS-N1 field classifications are presented in Figure 8 of \cite{Das2024}. For SFGs, $\sim$90\% of sources determined to be SFGs in \cite{Best2023} are also described as SFGs in the work of \cite{Das2024} however, this is closer to $\sim$70\% for the LERGs of \cite{Best2023}. As such, we acknowledge that differences in the classification process will affect {some of the samples of sources used} in this work. {We also note that recent work using physical processes to split sources by AGN and star formation physical processes using high resolution LOFAR data may indicate some underestimation of AGN activity in some sources \citep[see][]{Morabito2024}. }}

{In this work, we continue with the catalogues of \cite{Best2023}. This is because they are the source classifications which were used to study the luminosity functions of LOFAR detected sources \citep[][]{Kondapally2022, Cochrane2023}. The luminosity functions from these studies will be important to generate random catalogues which are necessary to measure the clustering of sources in this work, as described in Section \ref{sec:randoms}.  }

\section{Data and Random Catalogues for $\omega(\theta)$}
\label{sec:tpcf}

\subsection{Calculating $\omega(\theta)$ from Auto- and Cross-Correlations}
\label{sec:omega_eq}

As discussed in Section \ref{sec:intro}, the two-point correlation is {one commonly used method} {to study} the large-scale structure of galaxies within a {survey}. As the LoTSS sources are dominated by those with photometric redshifts we rely on the angular two-point correlation function, $\omega(\theta)$, to quantify the clustering within the fields. We measure $\omega(\theta)$ using the Landy-Szalay {estimator \cite{LandySzalay}}:

\begin{equation}
    \omega(\theta) = \frac{\overline{DD(\theta)} - 2\overline{DR(\theta)} + \overline{RR(\theta)}}{\overline{RR(\theta)}}.
    \label{eq:omega_auto}
\end{equation}

\noindent This uses normalised pairs of galaxies within the data catalogue, $\overline{DD(\theta)}$, {pairs in a} random catalogue, $\overline{RR(\theta)}$, and between the two catalogues, $\overline{DR(\theta)}$. {The normalisation ensures that across all $\theta$ bins, the sum of the normalised pairs sums to one, e.g. $\sum \overline{DD(\theta)} =1$.} The random catalogues should be a random distribution of galaxies, but that account for observational systematics in the data {and so mimic detection across the field of view}. Such systematics can be complex to account for \citep[see e.g. discussions in][]{Hale2024} and so we describe the creation of our random catalogues in detail in Section \ref{sec:randoms}.

{Whilst we can rely on the auto-correlation to measure {source clustering, in this work we also use the}} multi-wavelength data from \cite{Kondapally2021} and \cite{Duncan2021} to study the angular cross-correlation function between the radio and multi-wavelength {data}. {These multi-wavelength catalogues have a higher source density {than the} radio sources over the same area. {Cross-correlating between two catalogues can reduce the impact of any remaining systematics and help improve constraints on the {biases of the radio sources} by reducing the statistical uncertainties}. Combined, this can help {improve constraints on the} physical properties derived from modelling the angular clustering.} The angular cross-correlation function is given by:

\begin{equation}
    \omega_{CC}(\theta) = \frac{\overline{D_1D_2(\theta)} - \overline{D_1R_2(\theta)} - \overline{D_2R_1(\theta)} + \overline{R_1R_2(\theta)}}{\overline{R_1R_2(\theta)}}.
       \label{eq:omega_cross}
\end{equation}

\noindent Here ``1" and ``2" relate to the radio and the {multi-wavelength} catalogues respectively. Such a formalism has been used in a number of studies \citep[see e.g.][]{Hartley2013, Lindsay2014, Bielby2016, Krishnan2020}. 

\subsection{Random Catalogues}
\label{sec:randoms}

{As discussed, a} catalogue of randomly distributed sources is necessary to measure $\omega(\theta)$ {using both the auto- and cross-correlations. These random catalogues {must have no underlying large-scale structure, but must mimic the detection of sources across the fields,} accounting for observational effects and spatial masks. This means that the distribution of the random catalogue will be non-uniform. Observational effects, such as sensitivity variations, are more challenging to account for {and require understanding of the systematic effects which affect source detection}}. Therefore, either conservative flux density limits {should be} {applied or} these observational effects {need to be accounted for within the random {catalogues}. The latter approach allows more sources across the field to be used to measure the clustering for the population and so has been adopted in a number of studies, such as \cite{Hale2018, Mazumder2022, Hale2024}. In this work, we account for {the observational systematics {for our random catalogues and outline this process in the next sections. Figure} \ref{fig:randomsflowchart} provides a schematic representation of the steps involved. }

\begin{figure}
    \centering
    \includegraphics[width=1.02\linewidth]{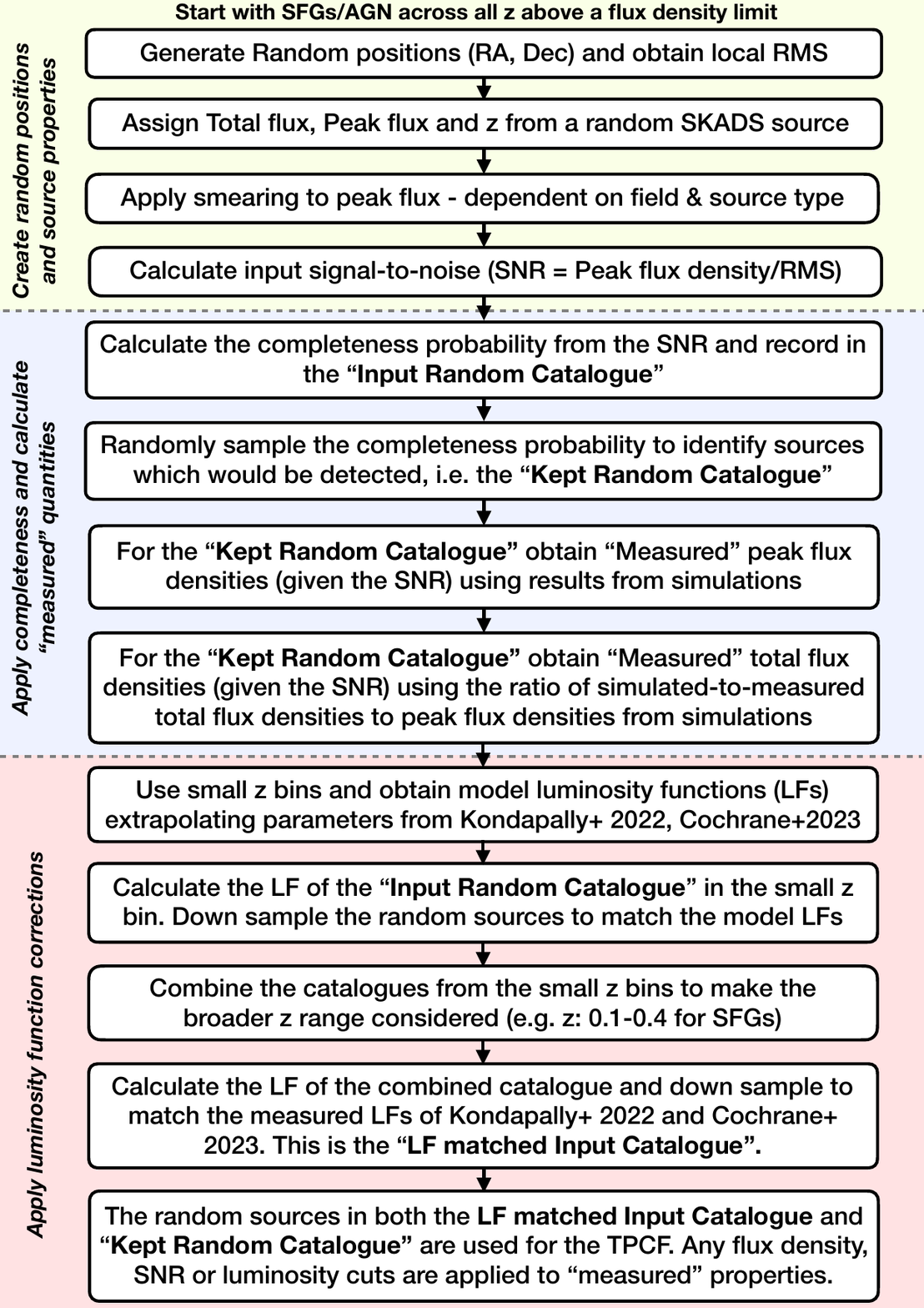}
    \caption{{Flowchart outlining the steps to make the catalogue of random sources associated with the radio data which are used to measure the clustering, divided into three stages. The first (yellow) relates to the creation of the general simulated source properties, as in Section \ref{sec:randoms_radio}. Second (in blue) describes the method of applying completeness effects and measurement errors, as in Section \ref{sec:randoms_radio}(i). Finally, is the effect of applying corrections for the intrinsic luminosity distribution (pink), as in Section \ref{sec:randoms_radio}(ii).} }
    \label{fig:randomsflowchart} 
\end{figure}

\subsubsection{Radio Random Catalogues}
\label{sec:randoms_radio}

To generate our {radio random catalogues we first generate positions across the LoTSS Deep Fields {over the regions that the radio data has been restricted to} \citep[{as outlined above,} namely the best ancillary regions of][]{Kondapally2021}. Each {position} is assigned {the source properties (flux density, redshift, shape) from a simulated source of} the SKA Design Studies \citep[SKADS;][]{Wilman2008, Wilman2010}, using the modified {SKADS} catalogue (with double the number of SFGs)} as used in \cite{Hale2024}\footnote{{This accounts for known differences between SKADS and faint source counts \protect \citep[see e.g.][]{Mandal2021, Hale2023}}}. A {peak flux density for the source is calculated by convolving the source model with the LOFAR 6\arcsec \ beam. We restrict the SKADS catalogue to integrated flux densities $S_{144 \textrm{MHz}}\geq$0.05 mJy \footnote{scaled from the 1.4 GHz flux densities in SKADS}} are used}. Whilst updated {radio simulations} are available from T-RECS \citep{Bonaldi2019, Bonaldi2023}, we found \citep[{similarly to}][]{Asorey2021} that the T-RECS source model for {bright AGN generated larger sources than anticipated, affecting source completeness.}\footnote{Though we note that a similar analysis {with T-RECS led to radio bias values broadly similar to} those derived using SKADS.} {Though AGN} are not the dominant source population {in this} work \citep[see][]{Best2023}, {it is important to consider such effects {and, as such, we used SKADS.}} \\

\noindent \textit{(i) Accounting for incompleteness and measurement errors}: \\

\noindent To generate the random catalogue, we {broadly followed the method of \cite{Hale2024}, who} used the results {from completeness simulations in the LOFAR Two-metre Sky Survey \citep[LoTSS-DR2;][]{Shimwell2022} to quantify (i) {completeness of} source detection as a function of input signal-to-noise (SNR); (ii) the measured-to-simulated peak flux density as a function of input SNR; and (iii) the ratio of the measured-to-simulated peak flux density compared to the measured-to-simulated integrated flux density as a function of SNR}. These factors {were combined with a distribution of sources from the modified SKADS catalogue described above to determine which sources would be considered detected within the data. \cite{Hale2024} also accounted for a positional dependent smearing of sources across the field of view}. {Combined, \cite{Hale2024} created a catalogue of random} sources which accounted for the detection across the field of view and had associated ``measured" peak and integrated flux densities. 

In this {work, we produce our own completeness simulations to be analogous to those of \cite{Shimwell2022}, using the methodology described in \cite{Hale2023}}. {This uses} an input source counts model (the modified SKADS catalogue, as above) at 144 MHz to generate simulated sources which are injected into the radio image. {We} then use the \texttt{PyBDSF} parameters of \cite{Sabater2021} and \cite{Tasse2021} to generate catalogues of sources which would be detected by \textsc{PyBDSF}. For each field 1000 simulations are run each with 2000 sources per simulation randomly injected into the image. These sources have a random flux density assigned from SKADS, with a source model that is convolved with the 6\arcsec \ LOFAR beam. For those sources that are detected by \textsc{PyBDSF}, they are matched to an input source using a 3.5\arcsec \ match radius. {This is smaller than the angular resolution to ensure these are true matches}. This output matched catalogue allows the calculation of the {the necessary measurements outlined above, such as the  completeness and measured source properties as a function of SNR. }

Using the catalogue of simulated random sources, we calculate their SNR based on their peak flux density and the rms at the source location. Using the results of the {completeness simulations and the methodology from \cite{Hale2024} to measure completeness as a function of SNR, we obtain the probability that each source in the catalogue of random sources is detected {and then use a process of random sampling to determine the random sources which will be used to measure the clustering}.} {Unlike in \cite{Hale2024}, we do not apply {position dependent smearing as we are unable to independently measure it and, in any case, smearing effects should be reduced given that only a smaller sky area closer to the pointing centre is used. Instead, {we apply a} constant smearing factor to the simulated peak flux density of the random sources in each field. This factor is allowed to be different for each sub-population in a field, as in addition to accounting for physical smearing effects, it can also empirically correct for differences between the simulated and true source size distributions. {These values are chosen to ensure} that the peak of the ratio of the {measured} integrated-to-peak flux density distribution for the simulated random sources matches that of the data within each field. {These factors varied in the} range of $\sim$1-1.15 across the fields.}} 

{{For those {random sources} which are considered detected, a ``measured" peak and integrated flux density} is then obtained based on the input SNR as in \cite{Hale2024}, using the {measured output-to-input} flux density distributions as a function of SNR found from the completeness simulations above.}  {These `measured' values are more similar to the flux densities in the \texttt{PyBDSF} data catalogues. Such flux densities have differences to the intrinsic flux densities due to both the} noise in the image and measurement differences introduced by the source finder. \\

{At this stage}, two catalogues of random sources are retained. The first contains the input {catalogue of random sources} {and their position,} {local} rms, simulated integrated and peak flux {densities} and the completeness probability for the {source. It} also contains a flag {for whether the source is considered to be {`detected' (or not)} from the completeness probability and {random sampling}}. We call this the {\textit{input random catalogue}}. The second catalogue contains {the subset of these sources which were considered} `detected' and for which  a ``measured" peak and integrated flux density are also recorded. We refer to this catalogue as the {\textit{`kept' catalogue of random sources}}. {The `kept' catalogue is the basis for the catalogue we use for the radio random terms in Equations \ref{eq:omega_auto} and \ref{eq:omega_cross} and will apply all SNR, spatial masks and flux cuts that are applied to the data to this catalogue of random sources (see Sections \ref{sec:mask} and \ref{sec:fluxSNRcuts}).} We ensure each catalogue of random sources is more numerous compared to the number of data sources to ensure that the errors will be dominated by uncertainties in the data. {The ratio of randoms to data is given in Table \ref{tab:bias_tab}.} \\

\noindent \textit{(ii) Ensuring an accurate intrinsic luminosity distribution for each population over {the z range}}: \\

\noindent Though the source counts distributions of the modified {SKADS catalogue} {agrees} well {with} deep radio surveys \citep[e.g.][]{Mandal2021, Matthews2021, vandervlugt2021, Hale2023}, we need to ensure that {that this remains true {when we split sources}} as a function of redshift, source type and flux density. To do this, we use {modelled} luminosity functions, $\Phi(L_{144 \textrm{MHz}})$\footnote{We will now drop the 144 MHz subscript when referring to luminosities such that $L$ refers to a spectral luminosity at 144 MHz, unless otherwise stated.}, of SFGs \citep{Cochrane2023} and the LERGs \citep[and QLERGs;][]{Kondapally2022}  in the LoTSS Deep Fields. We {use these models} to down-sample the {catalogue of random sources} so their input luminosity functions match the {models for each redshift bin/source type, as outlined below. }

{First, we use the 1/$V_{\textrm{max}}$ method \citep{Schmidt1968} to measure $\Phi(L)$ for the \textit{input catalogue of random sources}. This method is regularly used for radio luminosity functions} \citep[see e.g.][]{Mauch2003, Novak2017, Kondapally2022}. Following this {method $\Phi(L)$} is defined as:

\begin{equation}
\Phi(L) = \frac{1}{\Delta \textrm{log}_{10}(L)}  \frac{1}{A_{\textrm{corr}}}  \sum_i \frac{1}{V_{\textrm{max}}, i}, 
\label{eq:LF}
\end{equation}
 
 \noindent where: $\Delta \textrm{log}_{10}(L)$ is the width of the log luminosity bins used to {calculate} $\Phi(L)$; $A_{\textrm{corr}}$ is a correction for {the finite area of the observations}; and $V_{\textrm{max}, i}$ is the maximum {comoving} volume within which the $i^{\textrm{th}}$ source between $\textrm{log}_{10}(L)$ {and} $\textrm{log}_{10}(L)+ \Delta  \textrm{log}_{10}(L)$ can be observed within, given the sensitivity {of the data, and the redshift range being studied}\footnote{Though it can also be used, as in \cite{Novak2017}, to account for incompleteness effects within the data.}. {This work will study the evolving clustering of radio sources, using the redshift binning of \cite{Cochrane2023} for SFGs and \cite{Kondapally2022} for LERGs to $z\lesssim1.5$} \citep[where stellar masses are estimated to in][]{Duncan2021}. {In} the work of \cite{Cochrane2023}, {the lowest redshift bin considered {begins at} $z=0.1$.} Below such redshifts the {size of sources may} affect estimates of the {host galaxy properties}, using aperture {based} fluxes.

{Next, for an accurate input catalogue of random sources}, we {must} ensure that both the input luminosity distribution {and the redshift {distribution mimic} that of the data. This is to ensure an accurate flux density distribution {(or source counts)} within the redshift bin for the source type considered}. Therefore, we {use fine} redshift bins ($\Delta z=0.025$) to {match the input luminosity function of the catalogue of random sources to the model luminosity functions}. In each redshift bin, we {use a quadratic (linear) fit extrapolation of the} best fit parameters of the models from \cite{Cochrane2023} (or \cite{Kondapally2022}) to estimate {the luminosity function parameters (and therefore, models) within the $\Delta z=0.025$ bins. We compare $\Phi(L)$ for the input {catalogue of simulated} random sources assuming } {no incompleteness within the field and using a minimum flux density limit {(see Section \ref{sec:fluxSNRcuts})}} term\footnote{{{The input catalogue of random sources should have no incompleteness effects and be representative of the true underlying population}. The effects of incompleteness will be accounted for when a $\Phi(L)$ model matched sample of the catalogue of random sources is made and the sources which were considered `detected' in the catalogue generation earlier are used.}} in Equation \ref{eq:LF}\footnote{{In reality, $V_{\textrm{max}}$ should also account for limitations in the multi-wavelength catalogues. However, owing to the deep nature of the optical and IR data and the high fraction of host galaxy association (97\%), we neglect this compared to the $V_{\textrm{max}}$ of the radio emission.}}. {Using the ratio of the observed $\Phi(L)$ {for the input catalogue of random sources} to the model $\Phi_{\textrm{mod}}(L)$, we find the smallest value of this ratio across each $\Delta z$ and luminosity bin}\footnote{{As the ratio {in} the first and last luminosity bin may not be fully probed by the data or randoms, we do not use these values to find the minimum ratio.} The minimum ratio itself will be $\gg1$ due to the much higher number density of randoms.}. {We then normalise all luminosity functions of the input catalogue of random sources by this minimum ratio and downsample the random {sources to} match the luminosity function model in each $\Delta z$ bin.} {Combining the {random sources} from each of the $\Delta z$ bins} {in this way provides an input random catalogue with} a luminosity function that reflects the {intrinsic models and redshift distributions of radio sources in the Universe}. 

{However, we note that {the} parameterised models of the luminosity functions for the SFGs and LERGs {from} \cite{Cochrane2023} and \cite{Kondapally2022} are {smoothed models for $\Phi(L)$.} {In practice there may be larger} deviations between the model and the data than at {some} {luminosities}. This is more prevalent for {LERGs} \citep[see $0.5 <z \leq 1.0$ in Fig 6. of ][]{Kondapally2022}. To avoid large differences in the luminosity distributions of the {catalogue of random sources to the data {$\Phi(L)$, we downsample the input catalogues of random sources} across the full redshift bin range {to match} the measured luminosity functions of \cite{Kondapally2022} and \cite{Cochrane2023}.}} The random catalogues which are then used to {measure} $\omega(\theta)$ are the subset of this new input random catalogue that were determined to be {`detected'} in Section \ref{sec:randoms_radio}(i). The {sources in the `detected' catalogue} should then have luminosities, $z$ and flux density distributions which {are} similar to {those} of the observed data and {which also suffer from similar incompleteness effects across the fields}. \\ 

\subsubsection{{Multi-wavelength} Random Catalogues}
\label{sec:randoms_optIR}

As discussed in Section \ref{sec:omega_eq}, we also make use of the cross-correlation between the LoTSS Deep Fields data and the {multi-wavelength} catalogues within the field to trace the bias evolution of LOFAR {sources. This} requires an additional catalogue of random sources for the {multi-wavelength} catalogue {($R_2$ in Equation \ref{eq:omega_cross})}. For this we {use a} uniform distribution {of sources} across {the fields}. {This {assumes} that the mass and magnitude limits {applied to the {multi-wavelength data in Section \ref{sec:optIRdata}} provide high completeness and uniformity across each of the three fields.}}

\subsection{{Additional SNR and Flux density cuts}}
\label{sec:fluxSNRcuts}
As discussed in \cite{Hale2024}, the wavelet fitting mode which can be used with the source finding of \texttt{PyBDSF} can introduce the detection of a large number of sources below the nominal 5$\sigma$ detection {limit} across the rms {maps. Therefore, we} apply a 5$\sigma$ peak SNR cut to the radio catalogues for {both the data and catalogue of randoms}. Moreover, {we apply a constant flux density cut to normalise the different flux limits in the three fields}. {Since,} Bo\"otes is the shallowest field with a typical rms $\sim$ 44 $\muup$Jy beam$^{-1}$ over the multi-wavelength region \citep[see Tables in][]{Kondapally2021, Mandal2021}{, we} {therefore} impose a 200 $\muup$Jy {integrated} flux density limit {such that our data is at SNR$\gtrsim$5} in the shallowest field. 

{The catalogues of random sources {generated per field {are also} reduced in numbers to} ensure {a constant ratio of the number of data to randoms in each field and for each of the radio sources subsamples (e.g. split by redshift)}. This {avoids} spuriously large $\omega(\theta)$ (at higher $\theta$) {when a constant ratio was not applied}. For each {sub-sample} we ensure that the ratio of data to randoms is constant in each field and that this ratio is in the range of {$\sim$10-15, see} Table \ref{tab:bias_tab}.} \\

\noindent {At this stage we now} have the catalogues necessary to measure $\omega(\theta)$ across the combined three fields using both the auto-correlation (Equation \ref{eq:omega_auto}) and cross-correlation (Equation \ref{eq:omega_cross}).

\begin{figure*}
    \includegraphics[width=\textwidth]{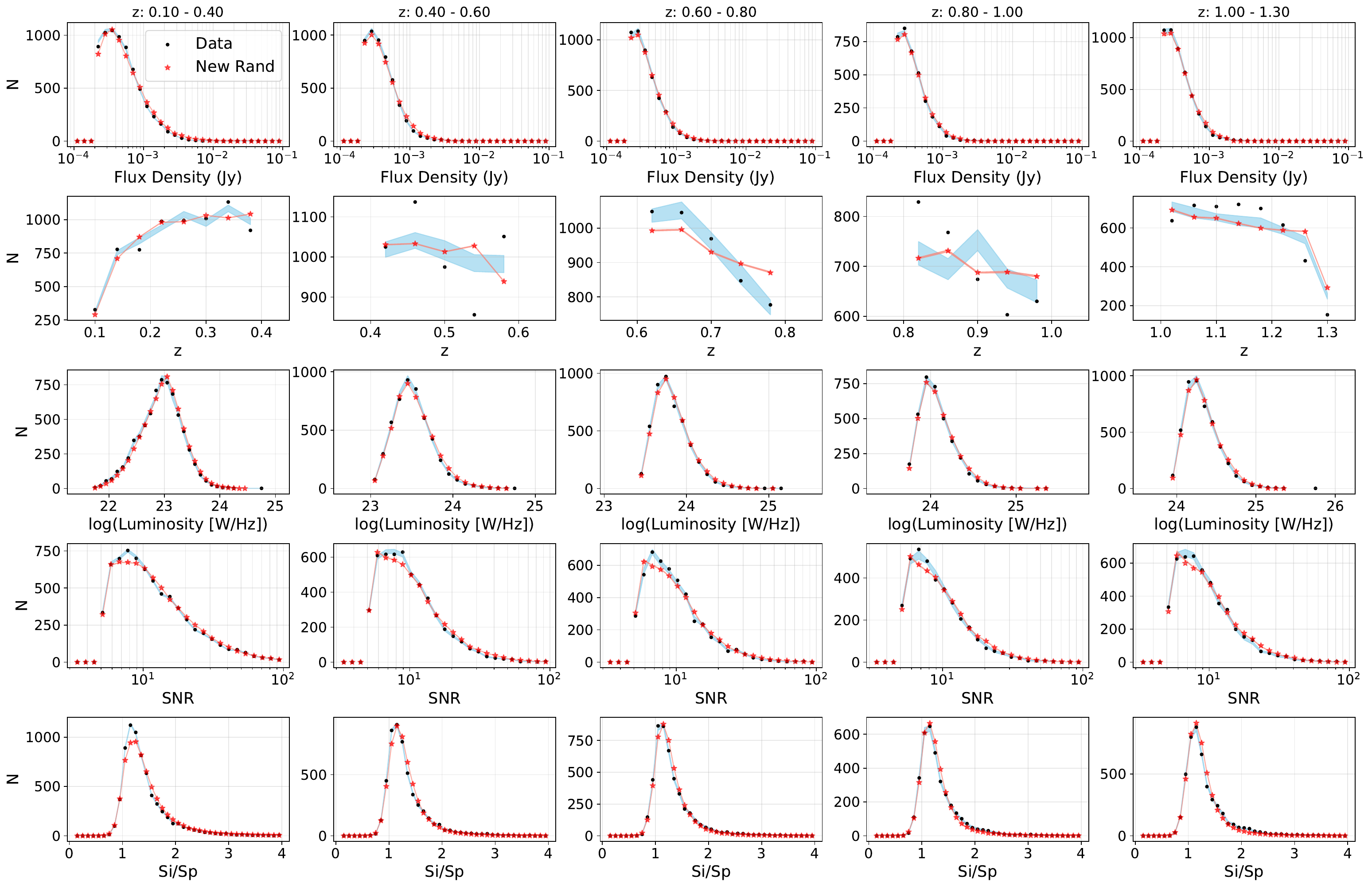}
    \caption{{Comparison plots of the flux density distributions (1st row); redshift distributions (2nd row); luminosity distribution (3rd row); signal-to-noise (SNR, 4th row); and integrated-to-peak flux density ratio ($S_I/S_P$; 5th row) for SFGs in the different redshift bins considered in this work, increasing in redshift from left to right. In each {panel}, the data catalogue with redshift cuts applied on the \texttt{Z\_BEST} column are shown as black dots, whilst each blue shaded {region represents} the output distribution from the data samples given in the range of the 16th - 84th percentiles of the values from the $p(z)$ resamples. The randoms for the full sample {are} shown as red stars. {These have associated red shaded regions with the range of randoms from those associated with each of the data $p(z)$ resample {(to ensure a constant ratio of random sources to data)}, though these are small as they are drawn from the same random sample and only have small differences reflecting the number of data per $p(z)$ sample.}} }
    \label{fig:flux_sfgs}
\end{figure*}

\begin{figure*}
    \begin{minipage}[c]{.54\textwidth}
        \includegraphics[width=8.5cm]{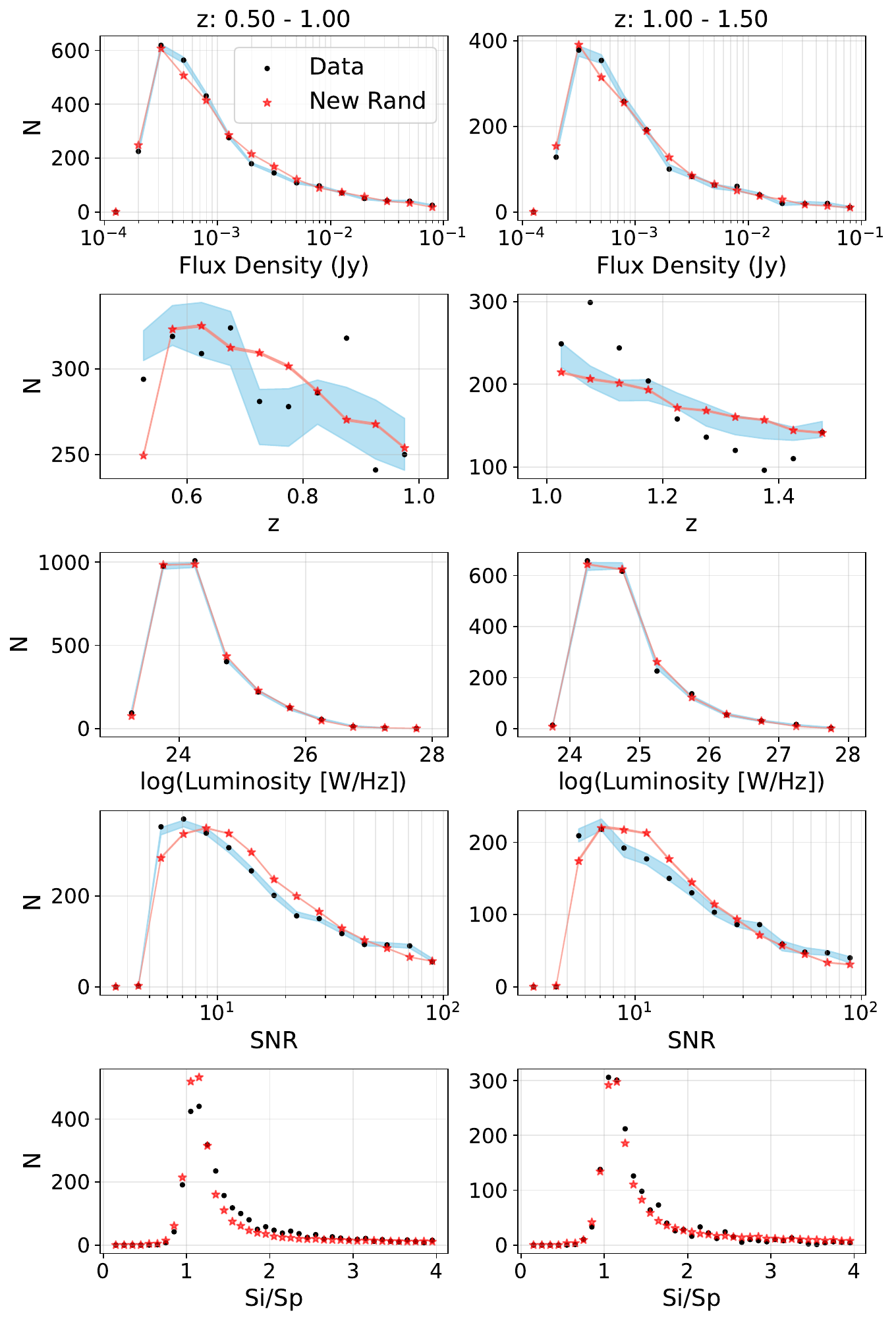}
        \subcaption{LERGs}
    \end{minipage}
    \begin{minipage}{.45\textwidth}
    \raggedright
        \includegraphics[width=8.5cm]{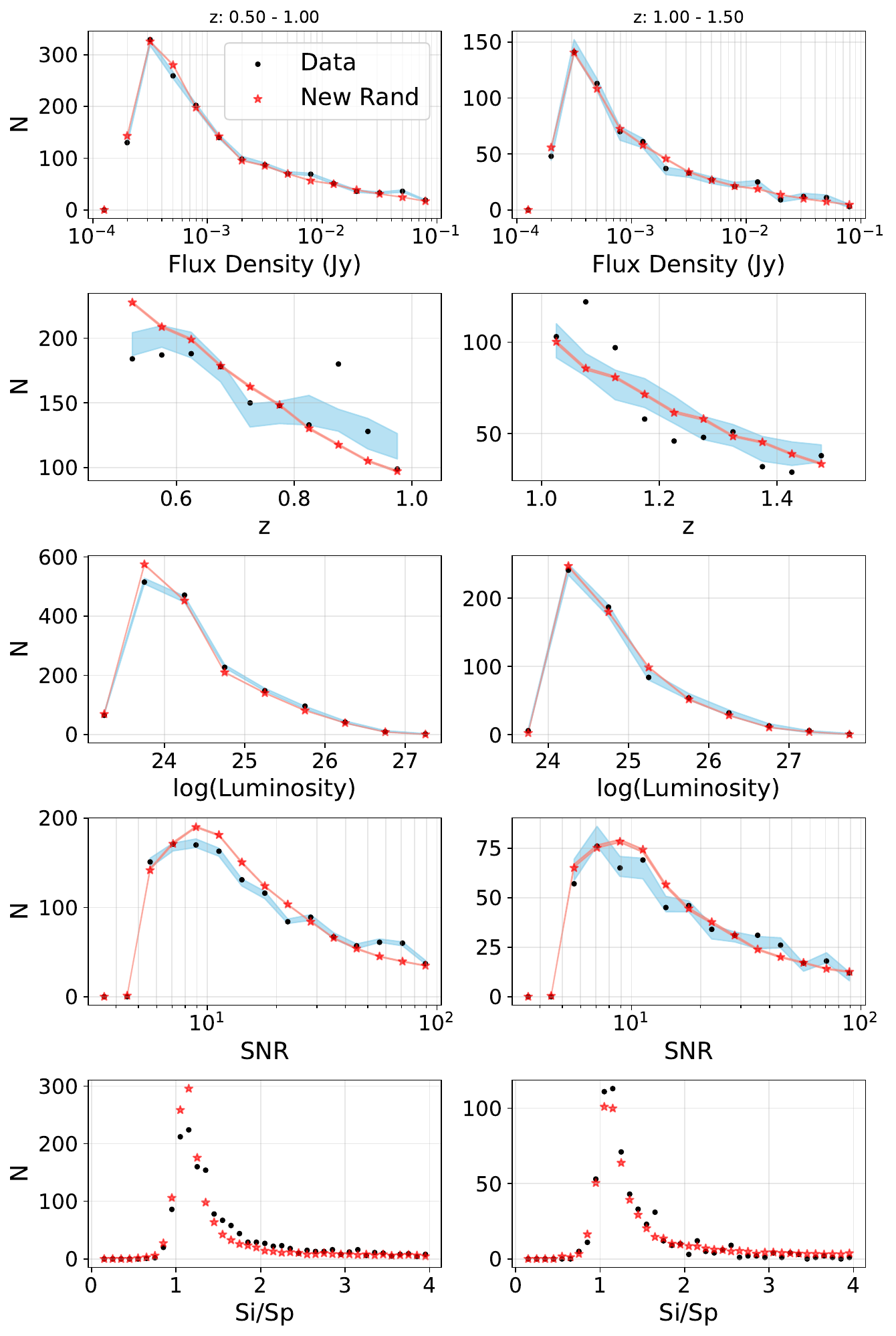}
        \subcaption{QLERGs}
    \end{minipage}  
    \caption{{As for Figure \protect \ref{fig:flux_sfgs} but for LERGs (left two panels) and QLERGs (right two panels). {Owing to differences in the source models between the data (where sources are assumed to be Gaussians) and the randoms (which are ellipses convolved with the beam) and that LERGs are likely to have more extended morphologies than for SFGs, we expect larger differences in $S_I/S_P$ for the LERGs than for SFGs compared to the randoms.}}}
    \label{fig:flux_lergs}
\end{figure*}

\subsection{Resampling of the data to probe $p(z)$}
\label{sec:pz_samps}
To determine the clustering as a function of redshift, and {accurate uncertainties on the measured clustering}, it is necessary to take account of the uncertainties in the redshifts of the sources, encoded in the redshift {posterior} probability distribution, $p(z)$, for each source. To do this, we construct {100} new redshift values for each source from sampling from the $p(z)$ derived {in the analysis of \cite{Duncan2021}}. For those sources with a spectroscopic redshift we use a constant value for $z$ in each resample. Combining the redshift resamples for all the sources provides {100} possible data samples for which we apply the necessary masking and flux density/SNR cuts and then compute the angular clustering for sources with a resampled redshift in the $z$ range being considered. 

\subsection{Comparison of Data and Random Catalogues}

Comparisons of the data to the randoms are presented in {Figure \ref{fig:flux_sfgs} for the SFGs and in Figure \ref{fig:flux_lergs} for the LERGs and QLERGs.} Shown are comparisons of the flux density, redshift, luminosity and SNR distributions for the randoms and data both when split into redshift bins using the \textrm{Z\_BEST} {redshift column} \citep[from the catalogue of][]{Duncan2021} and also using the resampled $z$ values from the $p(z)$. These allow us to demonstrate the accuracy of the random catalogues in accounting for the observational effects within the data.

As can be seen from Figures \ref{fig:flux_sfgs} and \ref{fig:flux_lergs}, the randoms broadly provide a good representation of the data, especially when the data from the resampled $p(z)$ {are} compared to. This suggests the random catalogues should provide a good simulated catalogue to measure $\omega(\theta)$.  {As discussed in \cite{Kondapally2021} and \cite{Cochrane2023}, at {some} redshifts, there can be large uncertainties on the photometric redshifts with some aliasing of the `\texttt{Z\_BEST}' value, whereas the $p(z)$ better captures this effect.} The redshift distributions do present some larger discrepancies within some of the sub-samples. However, we note that the flux density comparisons and SNR comparisons are the most important, as incompleteness relates to the {observed properties} of the source and knows nothing of their $z$ or luminosity. Provided these flux distributions appear appropriate, differences in the $z$ distribution are of less concern. Examples of such differences in the $z$ distribution can be seen for the $0.6 \leq z < 0.8$ redshift bin of SFGs, however the flux density and SNR distribution appear to be in good agreement {with} the data. For LERGs and QLERGs (see Figure \ref{fig:flux_lergs}), these distributions show {broad} agreement to those of the data resamples, though the differences in SNR distributions {are} greater than seen for SFGs. This is, in part, related to the source models used for such sources. As SKADS does not have L/HERG classification, we use a mixture of AGN \citep[Fanaroff Riley Type I and II sources][and radio quiet quasars]{Fanaroff1974}. However this may provide a mixture of source models not wholly {representative of} the demographics of LERGs.

For SFGs we also intend to study the luminosity dependence of the clustering of SFGs and so present similar plots to that as in Figure \ref{fig:flux_sfgs} for each of the luminosity ranges considered within each redshift bin investigated. These are presented in the Appendix in Figures \ref{fig:flux_sfgs_z0p1_0p4} - \ref{fig:flux_sfgs_z1p0_1p3} and again broadly show good agreement with the relevant data.

\section{$\omega(\theta)$ - Measurements, Results and Discussion}
\label{sec:omega}

To measure $\omega(\theta)$ we use \textsc{TreeCorr} \citep{Treecorr} to calculate the pairs of galaxies within different angular separation {bins from our data and random catalogues} {and then use these alongside Equations \ref{eq:omega_auto} and \ref{eq:omega_cross} to measure the auto- and cross- angular correlation functions}. {Aside from differences in the angular bins used, we adopt} the same parameters for \textsc{TreeCorr} as in \cite{Hale2024} and subsequently use these pairs to calculate $\omega(\theta)$ as in Equations \ref{eq:omega_auto} and \ref{eq:omega_cross}, ensuring to correctly normalise for the number of possible pairs across full angular range. In order to determine the {impact of the redshift (and its uncertainties) on our clustering measurements}, we calculate {$\omega(\theta)$} for both those sources split by \texttt{Z\_BEST} and those split into the redshift bins using the 100 resampled redshifts for each source. The $\omega(\theta)$ presented for the $p(z)$ resampled data is the mean $\omega(\theta)$ from the resamples in each angular bin. 

We estimate the uncertainties on $\omega(\theta)$ using a process of {bootstrapping \citep[see e.g.][]{Ling1986} where we resample the data to generate new samples of the data which have the same size as the original data sample but containing randomly selected sources and then using these to calculate $\omega(\theta)$.}\footnote{{We note that other methods to generate errors are possible such as Jackknife errors and using bootstrapping with sub-regions as opposed to individual sources. We choose to remove individual sources, which more closely mimics how we resample galaxies based on their $p(z)$. Whilst this can be found to underestimate uncertainties in some conditions \citep[see e.g.][]{Norberg2009}, bootstrap resampling using sub-volumes can also be found to overestimate errors. We take this individual source approach for more consistency with previous work of \citep{Hale2018} and note {that we will discuss these errors in Section \ref{sec:results_bz}}.}} As such, a given source may be repeated in a sample or may be missing from a given bootstrap resample. We repeat this process 100 times. Furthermore, when we consider the effect of the $p(z)$ resampling, we estimate the error by repeating the bootstrapping process for each of the resampled $p(z)$ data sets. The errors are calculated from the combination of all of the bootstrap resamples in each of the redshift resamples and determining the uncertainties {as would be measured for a set of bootstrap resamples \citep[as described in][]{Norberg2009}. }

\begin{figure*}
    \centering
    \includegraphics[width=\textwidth]{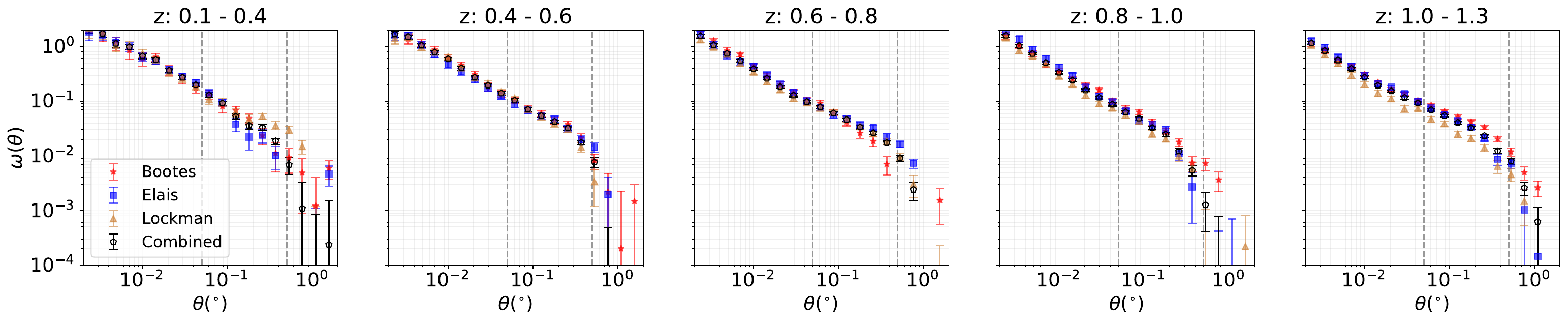}
    \caption{{{Auto correlation of the {multi-wavelength} sample of galaxies in each of the redshifts bins used to measure $\omega(\theta)$ for SFGs. Black open pentagons indicate the} {combined TPCF} {across the three fields, with their individual $\omega(\theta)$ shown for Bo\"otes (red stars), ELAIS-N1 (blue squares) and Lockman Hole (gold triangles). {The dashed vertical lines highlight the region used to fit the correlation function over in order to measure the bias.}}}}
    \label{fig:tpcf_optIR_sfg}
\end{figure*}

\begin{figure}
    \centering
    \includegraphics[width=0.48\textwidth]{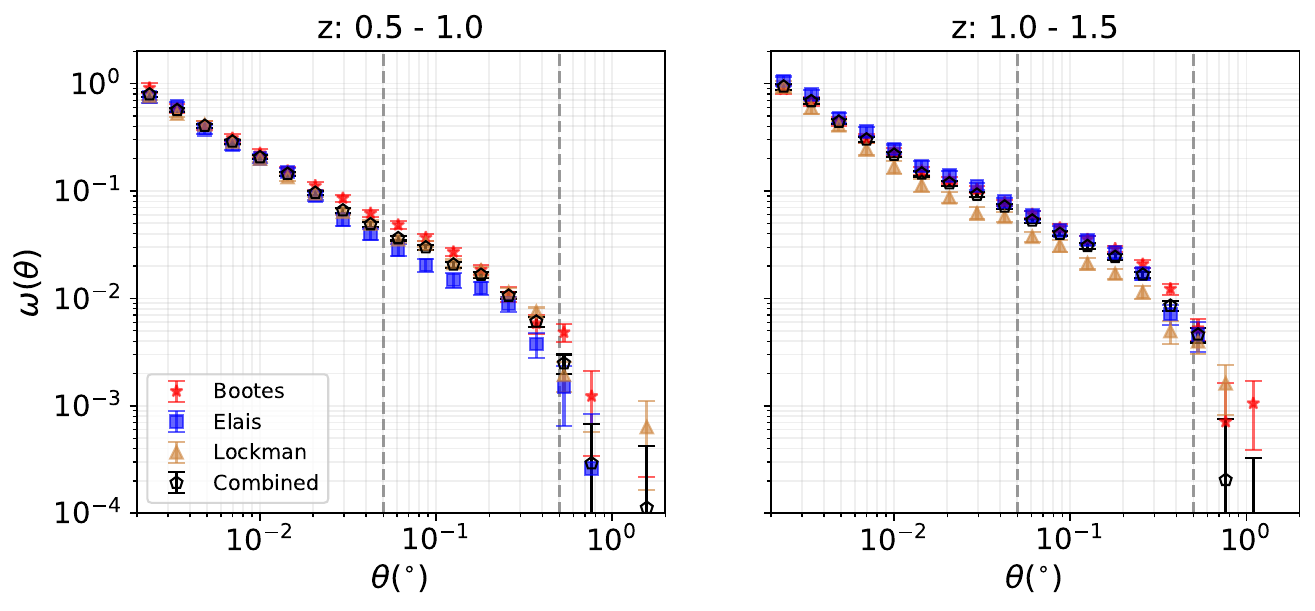}
    \caption{{{Similar to} Figure \protect \ref{fig:tpcf_optIR_sfg}, {shown are the auto-correlation for the multi-wavelength sample within the redshift bins used for the LERG (and QLERG) studies}.}}
    \label{fig:tpcf_optIR_lerg}
\end{figure}

{We only resample redshifts for the radio catalogue, not for that of the optical catalogue}. Whilst there are also uncertainties in the redshift distributions of the {multi-wavelength} catalogues, these galaxies are used as a reference sample in the cross-correlation. {As the properties of the galaxies such as their mass (which are used in the completeness cuts) are calculated assuming the measured `best' redshift of the sources, re-calculating such mass parameters for a different redshift is an intensive process and so it would be challenging to implement such a resampling of redshifts for the {multi-wavelength} galaxies. We therefore make redshift cuts for the {multi-wavelength} galaxies on the `best' redshift for the source.} The auto-correlation functions for the {multi-wavelength} data {are} shown in Figures \ref{fig:tpcf_optIR_sfg} and \ref{fig:tpcf_optIR_lerg} for the $z$ binning of the SFGs and (Q)LERGs respectively.

\begin{figure*}
    \centering
    \includegraphics[width=\textwidth]{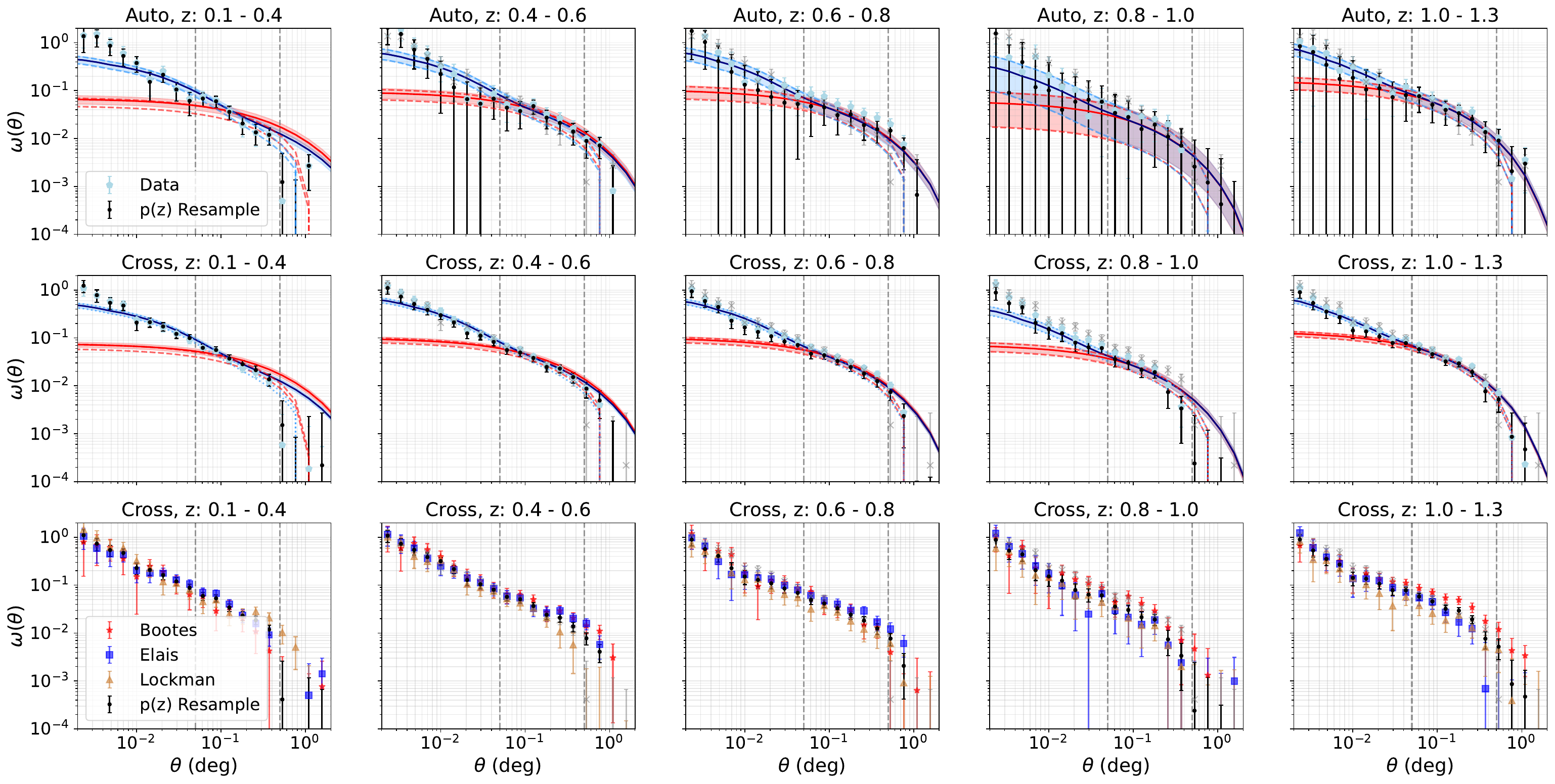}
    \caption{{Comparison of $\omega(\theta)$ for SFGs in different $z$ bins (left to right), using $z$ cuts based on (i) the data's best redshift value in the source catalogue (light blue) and (ii) from the $p(z)$ resampled value of $z$ (black). This is shown in increasing redshift bins (left to right) and for the auto- (upper panel), cross-correlation (middle panel) and for the cross-correlation compared to each of the three individual fields (lower panel) for Bo\"otes (red stars), ELAIS-N1 (blue squares) and Lockman (gold triangles). In the {top and middle panels} the red line and shaded region represents the best fit `Linear' model, whilst the blue line the best fit `HaloFit' model to the black data points. {Also shown is the model minus the integral constraint indicated by the red dashed line (for the `Linear' model) and the blue dotted line (for the `HaloFit' model).} {Dashed vertical lines indicate the $\theta$ ranges which we fit the data over. The grey crosses indicate the value of the $p(z)$ resampled $\omega(\theta)$ for the combined fields in the lowest redshift bin, purely to guide the eye. }}}
    \label{fig:tpcf_sfgs}
\end{figure*}

\begin{figure}
    \centering
    \includegraphics[width=0.5\textwidth]{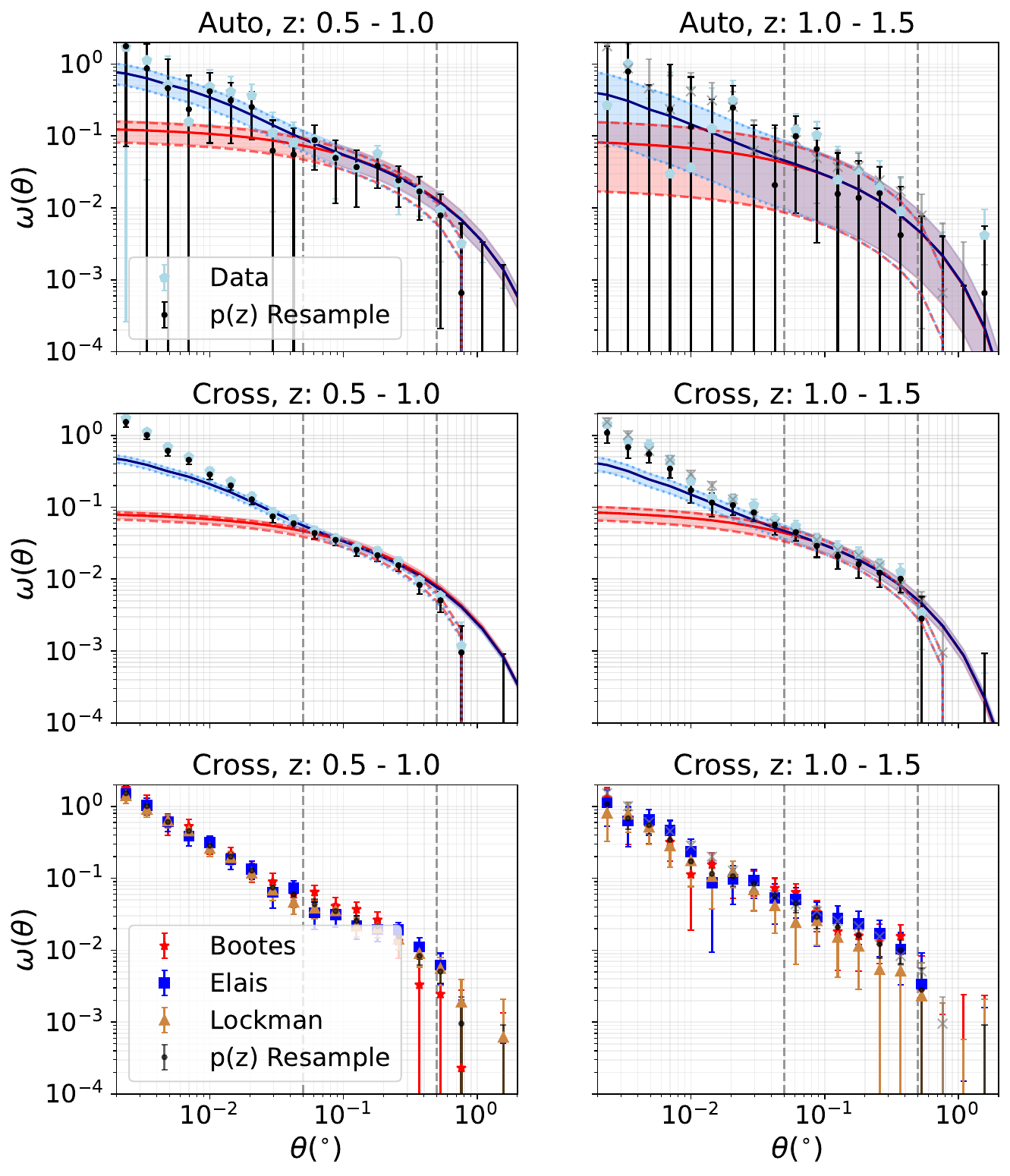}
    \caption{{As for \protect Figure \ref{fig:tpcf_sfgs} for the LERG {samples}.}}
    \label{fig:tpcf_lergs}
\end{figure}

\begin{figure}
    \centering
    \includegraphics[width=0.5\textwidth]{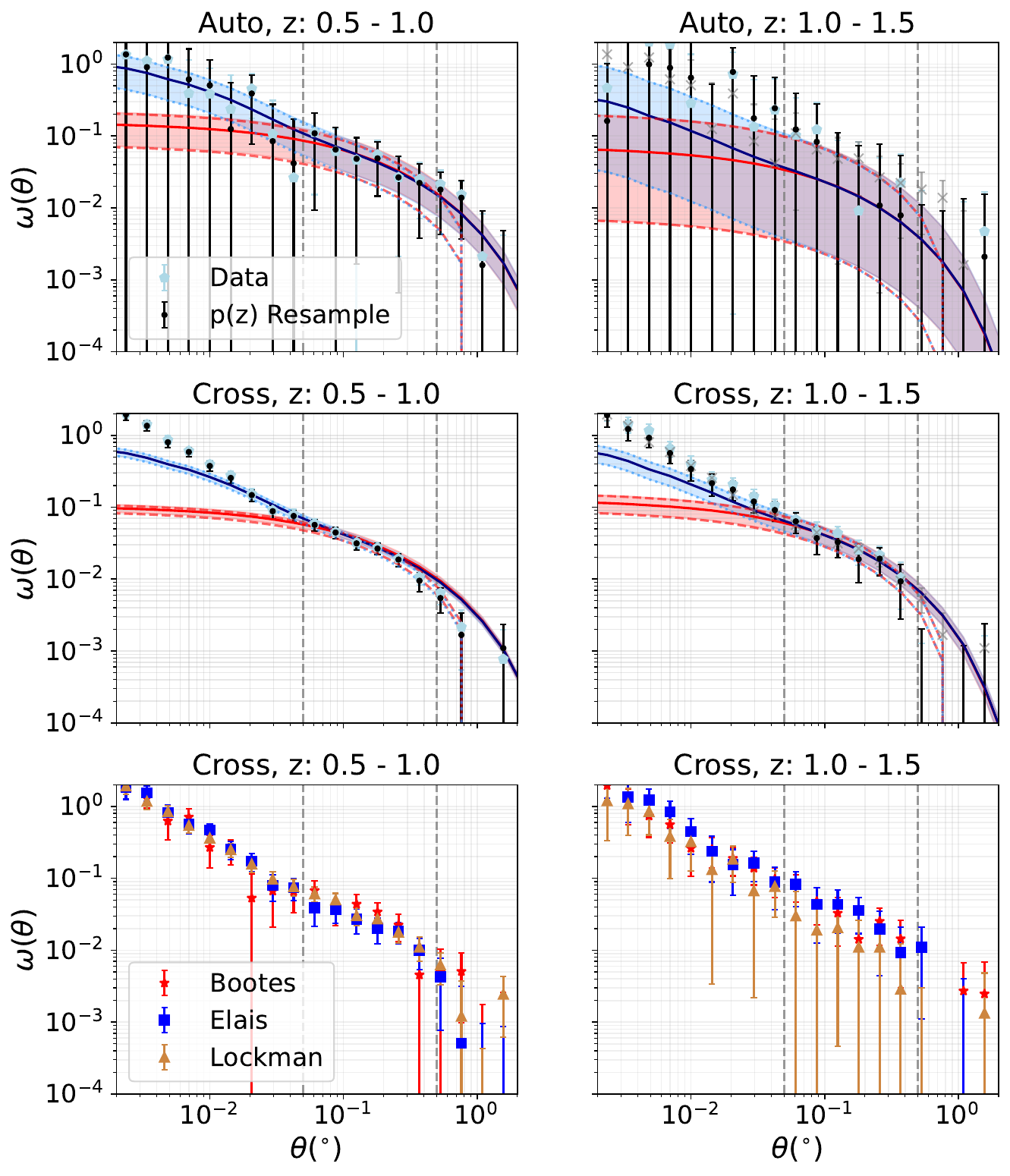}
    \caption{{As for \protect Figure \ref{fig:tpcf_sfgs} for the QLERG {samples}.}}
    \label{fig:tpcf_qlergs}
\end{figure}

We present a comparison of both the auto- and cross-correlation functions for SFGs, LERGs and QLERGs in Figures \ref{fig:tpcf_sfgs}, \ref{fig:tpcf_lergs} and \ref{fig:tpcf_qlergs} respectively. This is shown for both the redshift binning based on the separation using the \textrm{Z\_BEST} column the catalogues of \cite{Kondapally2022} and \cite{Duncan2021} and also from the resamples which probe the full $p(z)$. These figures demonstrate that the measured clustering within the fields can exhibit differences when the full $p(z)$ is not used to associate a redshift, which in turn will affect measurement of bias, though we note that most differences are within the uncertainties of the measurements. More broadly, it can be seen that these measurements of $\omega(\theta)$ from both the auto- and cross-correlations exhibit close to expected power law behaviour on the majority of angular scales, up to $\sim$0.5\degree. {At larger angular} scales, the clustering {signal} declines more sharply, {this is in part as} a result of the finite field size limiting the number of observable pairs of galaxies {at the largest angular scales}. {When fitting our model we will account for this using} an integral constraint \citep[see e.g.][]{Roche1999} {evaluated across the size of individual fields}. 

{At the smallest angular scales there is continued increased clustering to the smallest scales considered. In shallower radio surveys with comparatively more AGN}, such {clustering} at small angular scales often has a significant contribution from the clustering between multiple components associated with a single galaxy which have not {been} combined together into a single source \citep[see e.g.][]{Blake2002,Overzier2003}. However, as radio components have already been cross-matched in the work of \cite{Kondapally2021}, we are likely observing {genuine departures} from the large-scale power-law like clustering due to the `1-halo' clustering {- i.e.} the clustering of sources within the same dark matter halo \citep[see e.g.][]{Kravtsov2004, Zehavi2004}. 

Fitting the 1-halo clustering is possible within a {Halo Occupation Distribution (HOD) framework. Such a method allows the properties of the haloes that can host both central and satellite galaxies to be measured, under the assumption of an  HOD parameterisation} \citep[see e.g.][]{Berlind2002, Zheng2005, Zheng2007, Hatfield2016}. Such modelling of the full HOD parameterisation is beyond the scope of this work, but will be considered in future work, with deep surveys from telescopes such as LOFAR and {MeerKAT \citep[e.g.][]{Hale2024b}. We instead focus on the larger scale, 2-halo clustering, in order to measure the galaxy bias and how this evolves with redshift for different source populations}. {However, we will present our results of fitting $\omega(\theta)$ with models from the cosmology code the Core Cosmology Library \citep[\texttt{CCL},][]{Chisari2019} which take into account both the (i) 2-halo clustering only (the `linear' model) and (ii) a model which combines the 2- and 1-halo clustering \citep[the `HaloFit' model, see e.g.][]{Smith2003, Takahashi2012} in Section \ref{sec:results_bz}. {This is under default HOD used in \texttt{CCL} for the model, which may not be appropriate for the radio sources, especially on the smallest angular scales}. Both models are fit to the data to demonstrate that irrespective of the model assumed, we measure comparable values for the large-scale bias. }

As can be seen from Figures \ref{fig:tpcf_sfgs} - \ref{fig:tpcf_qlergs}, at low redshifts the difference in both the auto- and cross-correlations between the values of $\omega(\theta)$ when using the \texttt{Z\_BEST} values and the $p(z)$ resamples are small, likely owing to the fact that spectroscopic redshifts will likely dominate at low redshifts and sources are more likely detected across a wealth of multi-wavelength bands. Therefore, the differences between the $p(z)$ samples and the  \texttt{Z\_BEST}  selected sample is reduced, compared to higher redshifts. Figures \ref{fig:tpcf_sfgs} - \ref{fig:tpcf_qlergs} also demonstrate the large uncertainties found in the auto-correlation, especially for SFGs and LERGs at high redshifts, are reduced when the cross-correlation is instead measured. This is especially true for the LERGs, where the reduced number of sources compared to the {SFGs presents} challenges in measuring the bias from the auto-correlation function alone. We will therefore present the comparison of the bias measurements from the auto-correlation for the SFGs in Section \ref{sec:results_bz} and then proceed with the cross-correlation functions to measure the bias evolution of {SFGs and LERGs} in the LoTSS Deep Fields to draw conclusions.

\section{Galaxy bias Results and Discussions}
\label{sec:bias}

\subsection{Measurement of Galaxy bias, $b$}
\label{sec:meas_bias}

To measure the bias from $\omega(\theta)$, we follow the methodology in \cite{Hale2024} and {use \texttt{CCL}, which} uses cosmology packages such as \texttt{CAMB} \citep{CAMB} and \texttt{CLASS} \citep{CLASS} to generate models of the power spectrum and infer the angular clustering, through assuming cosmological parameters, bias models and redshift distributions. As in \cite{Alonso2021} and \cite{Hale2024} we use an evolving galaxy bias model \citep[$b(z) = \frac{b_0}{D(z)}$, where $D(z)$ is the growth factor, see e.g.][]{Hamilton2002} to quantify the {evolution of the} bias for a given population of sources {within the redshift bin}. A constant bias model was also used in {\cite{Alonso2021}, \cite{Hale2024} and \cite{Nakoneczny2024}}, but we note that due to the narrow redshift ranges considered {in this work,} the bias measured assuming a constant bias model ($b(z) = b$) showed little {{differences compared to} when the evolving galaxy bias model was assumed {when evaluated at the average redshift in the bin being considered}}.

In order to measure galaxy bias, a redshift distribution is also required for the radio and optical sources. {For the radio sources we use a different redshift distribution for each of the $p(z)$ resamples. This is taken the combined histogram of the resampled redshifts for each source in the resample, normalized to form a $p(z)$. Using the resampled $z$ values avoids unphysical spikes in the combined redshift distribution for all sources in the sample, which would be produced from spectroscopic redshifts. For the {multi-wavelength} sources we take a similar approach to create a combined redshift distribution for sources with a \texttt{Z\_BEST} value within the redshift range. These redshift distributions are therefore peaked within the redshift bin, but with wings in the $p(z)$ to redshifts beyond the bin value, due to uncertainties in the redshift values. We discuss this further in Section \ref{sec:limitations}.} 

The redshift distribution is provided to \texttt{CCL} and $b_0$ is determined for the auto correlation through firstly calculating the \texttt{CCL} model assuming $b_0$=1 and then scaling by $b^2$. This allows the $\chi^2$ distribution to be calculated as a function of bias, {using the full covariance as given by:}

\begin{equation}
    {\chi^2 = (\overrightarrow{\omega} - \overrightarrow{\omega_M})^T \textrm{Cov}^{-1} (\overrightarrow{\omega} - \overrightarrow{\omega_M}),}
    \label{eq:chi_cov}
\end{equation}

\noindent {where $\overrightarrow{\omega}$ is the angular two-point correlation function which is measured for the data sources, $\overrightarrow{\omega_M}$ is the modelled $\omega(\theta)$ which includes the subtraction of an integral constraint \citep[see e.g.][to account for the limited field sizes]{Roche1999} and $\textrm{Cov}$ is the covariance matrix calculated from the bootstrap resampling methods across the $p(z)$ samples considered. The covariance takes into account the correlations between $\theta$ bins which may impact the inferred bias values, compared to when the {diagonal elements alone (i.e. the errorbars in Figures \ref{fig:tpcf_optIR_sfg}-\ref{fig:tpcf_qlergs}) are used.}}

In the work of \cite{Hale2024} both the linear and HaloFit models \citep{Smith2003, Takahashi2012} within \texttt{CCL} were used to model the angular clustering. \cite{Hale2024} determined that the linear model was more appropriate for the LoTSS-DR2 data across the angular ranges considered, where {data at} $\theta \leq 0.03^{\circ}$ could also not be used to fit the bias in \cite{Hale2024} due to the excess clustering at small angular scales being partly attributable to multi-component sources. Due to differences in the `linear' and `HaloFit' models, these were in the best agreement when fitting above $\sim 0.3^{\circ}$. Due to the smaller maximum angular separations which can be probed in this work, we must use different $\theta$ ranges to fit the data, where we use 0.05\degree \ $\leq \theta <$ 0.5\degree.

{We fit for both the `HaloFit' and `linear' models and fit for $b_0$ through minimising {$\chi^2$}. We determine the uncertainties on $b_0$ through modelling the probability distribution from the $\chi^2$ distribution of $b_0$ (assuming $P\propto e^{-\chi^2/2}$). We randomly sample from this distribution and use this to determine the associated median, 16th and 84th percentiles for $b_0$. {To account for uncertainties introduced due to the $p(z)$ distribution of sources we fit the galaxy {bias}} for each $\omega(\theta)$ measured for the $p(z)$ subsamples. Combining together the randomly sampled bias values from fitting each of these $\omega(\theta)$ then gives a larger sample of bias values which we use to then quote the associated median and errors from the 16th and 84th percentiles. }

To determine $b_0$ from the cross-correlation, $\omega_{CC}(\theta)$, we follow a similar method to that for the auto-correlation, but using two tracers. This makes use of the relationship:

\begin{equation}
    b_{CC}^2 = b_{AC, 1} b_{AC, 2}
\end{equation}

\noindent where $b_{AC, 1}$ is the bias of the first sample (radio) and $b_{AC, 2}$ is the bias for the second sample ({multi-wavelength}), {as also used in works such as \cite{Lindsay2014}\footnote{{Whilst \protect \cite{Lindsay2014} include a growth factor term in their work, as we fold in the evolving bias and redshift distribution of the two populations in the modelling in \texttt{CCL} and do not evaluate at a single average redshift, this is not {believed to be necessary for this work}.}}}. $b_{AC, 2}$ is determined from the auto-correlation of the {multi-wavelength} data alone, with the redshift distribution {of the multi-wavelengths sources taken {as the combined $p(z)$ of sources with \texttt{Z\_BEST} values within the given redshift range.} To determine the bias of the radio sample from the cross-correlation we calculate $\omega_{CC}(\theta)$ assuming the {radio bias,} $b_{0, \textrm{radio}}=1$. We then follow a similar method to the auto-correlation and, {in every redshift bin for the population being considered,} scale this correlation function using the radio bias {(having assumed an optical bias, discussed below)}. However, in contrast to the auto-correlation, we now scale by the radio bias, $b$, as opposed to $b^2$. {{In this way, by varying $b$} and scaling the cross correlation function by this, we are able to again {measure} the probability distribution of bias and quantify the best fit of $b$ for each radio source population in the given redshift bin.}

Uncertainties on the radio bias from the cross-correlation need to account for both uncertainties in the measured values of $\omega_{CC}(\theta)$ {(which include the uncertainties in the {$p(z)$ of the radio sample})} and the uncertainties in the bias of the {multi-wavelength} sources. {We therefore, calculate the {radio bias, $b$,} through drawing 100 random samples {of the bias from auto-correlation of the optical sample}. For each optical bias value we combine this with the $\omega_{CC}(\theta)$ {from} the $p(z)$ resampling and use this to calculate the {radio bias through evaluating the $\chi^2$ and solving similarly to the auto-correlation.} After combining the radio bias samples derived for each of the resamples we have a bias distribution for the radio sample which is derived from the cross-correlation and accounts for the redshift uncertainties in the radio sources and uncertainties in the {multi-wavelength} bias. The bias values reported are then taken as the median bias values and associated errors are calculated from the 16th and 84th percentiles.} {We will present a comparison of the bias results for the Linear and HaloFit models respectively in Section \ref{sec:results_bz}, to demonstrate the effect it has on our measurements of $b$. The properties of the data in the redshift bins considered and the bias fitting parameters (assuming the Halofit model) are presented in Table \ref{tab:bias_tab}.}

\subsection{$b(z)$ Results for SFGs vs. LERGs}
\label{sec:results_bz}
We present our measurements of bias for SFGs and LERGs in Table {\ref{tab:bias_tab} and in} Figures \ref{fig:bias_z} and \ref{fig:bias_z2} alongside the comparison to previous models {{adopted} in \cite{Wilman2008, Wilman2010} and for the {previous} measurements of \cite{Nusser2015} and \cite{Lindsay2014} which are flux-limited samples (dominated by AGN), and for the classified samples (AGN vs. SFG) of \cite{Magliocchetti2017, Hale2018, Chakraborty2020} and \cite{Mazumder2022}.} As discussed, \cite{Hale2018}, use the VLA 3 GHz COSMOS Survey \citep{Smolcic2017, Smolcic2017b} to study the clustering of SFGs and AGN, as well as high redshift analogues for HERG and LERG populations. This provides the closest comparison to the studies presented in this work. {However, the {classification} adopted for the clustering of LERGs in \cite{Hale2018} is more {similar to the quiescent LERG (QLERG) population discussed in \cite{Kondapally2021} and adopted in this work}. {In the SKADS models, fixed halo masses were assumed for each population using the formalism of \cite{Mo1996} and we highlight these halo masses on Figure  \ref{fig:bias_z2}}.} {We note, though, that the masses assumed by \cite{Wilman2008} will not be a directly transferable to the full population of sources considered as this make assumptions about the source populations being dominated by central (not satellite) galaxies, see e.g. the works of \cite{Aird2021} and so we use them indicative only for comparisons.} 

Figure \ref{fig:bias_z} presents the bias measured from the auto-correlation and cross-correlation, for both the `linear' and `HaloFit' derived models. {We find} good agreement, in general, between the auto-correlation and cross-correlation methods, which are consistent within 1$\sigma$, as well as good agreement when the `linear' and `HaloFit' models are compared. This provides confidence that the measured bias values are not being affected by the choice of model. {We note, though, that constraints on the auto-correlations can be very uncertain, which is evident to be the case considering the auto-correlation in Figures \ref{fig:tpcf_sfgs} - \ref{fig:tpcf_qlergs}}. {We also note that when considering the fitting of $\omega(\theta)$, the minimum reduced $\chi^2$ (hereafter R-$\chi^2$) {values found can be} $\ll 1$ ({where a value of 1} would be expected for a good fit of the data), suggesting that our estimation of the uncertainties in $\omega(\theta)$ may be larger than {they should be}. {We note though that the R-$\chi^2$ of all resamples which are generated (from which the bias is obtained from the 16th, 50th and 84th percentiles) will have larger average R-$\chi^2$, as these values in the Table represent the minimum possible R-$\chi^2$ found.} As discussed, we have aimed to combine uncertainties on the TPCF (through bootstrap resampling), cosmic variance (through combining the three deep fields) and uncertainties in the redshift distributions of our radio sources {(through the $p(z)$ resamples)}. {{The R-$\chi^2$ found in the fitting of $b$ could therefore be indicative that we have provided too conservative values for the uncertainties in $\omega(\theta)$, which have folded through to the fitting of $b$}. This therefore could suggest that either (i) {the variance between fields is larger than is expected}, (ii) that the uncertainties associated with the $p(z)$ for the sources may be too broad for a subset of sources or this is related to the uncertainty method used, or (iii) that the spread between the fields is a result of remaining systematics or classification issues per field.  Uncertainties in the redshift distribution will be greatly reduced with the upcoming WEAVE-LOFAR \citep{Smith2016} survey, which will provide spectroscopic follow up of LOFAR detected sources, thus accurately constraining redshifts for a significant population of sources, and allowing for direct spatial clustering measurements as well as aiding in the source classification process}.  } 

In general, the constraints on {$\omega(\theta)$} from the cross-correlation are {less uncertain} than from the auto-correlation alone. The comparisons in Figure \ref{fig:bias_z2} between the three populations also shows, in agreement with \cite{Hale2018}, that over medium redshifts ($z\sim 0.5-1.0$) LERGs and QLERGs appear to be more biased tracers of {dark matter compared to those radio sources classified as SFGs}. {Comparing to those models assumed in SKADS, this supports} the idea {from other radio clustering studies} that AGN are typically found in more massive haloes than star formation dominated radio sources \citep[see e.g.][]{Magliocchetti2017, Hale2018, Mazumder2022}. This {is also in line with} numerous studies at other wavelengths and in simulations where redder galaxies are typically more clustered than blue galaxies \citep[see e.g.][]{Somerville2001, Zehavi2005, Coil2008, Cresswell2009} and may reflect the LERGs residing in galaxies {that typically have larger stellar masses than SFGs \citep[which can be demonstrated from the average consensus masses of][for which the median value is given in Table \ref{tab:bias_tab}]{Best2023}. We note though that at the highest redshift bin for the SFG population, the average mass is similar to that for the LERGs}. These in turn may be hosted by more massive haloes, given correlations between galaxy clustering and stellar mass \citep[see e.g.][]{Farrow2015, Cochrane2017, Durkalec2018}. 

\begin{figure*}
    \centering
    \includegraphics[width=\textwidth]{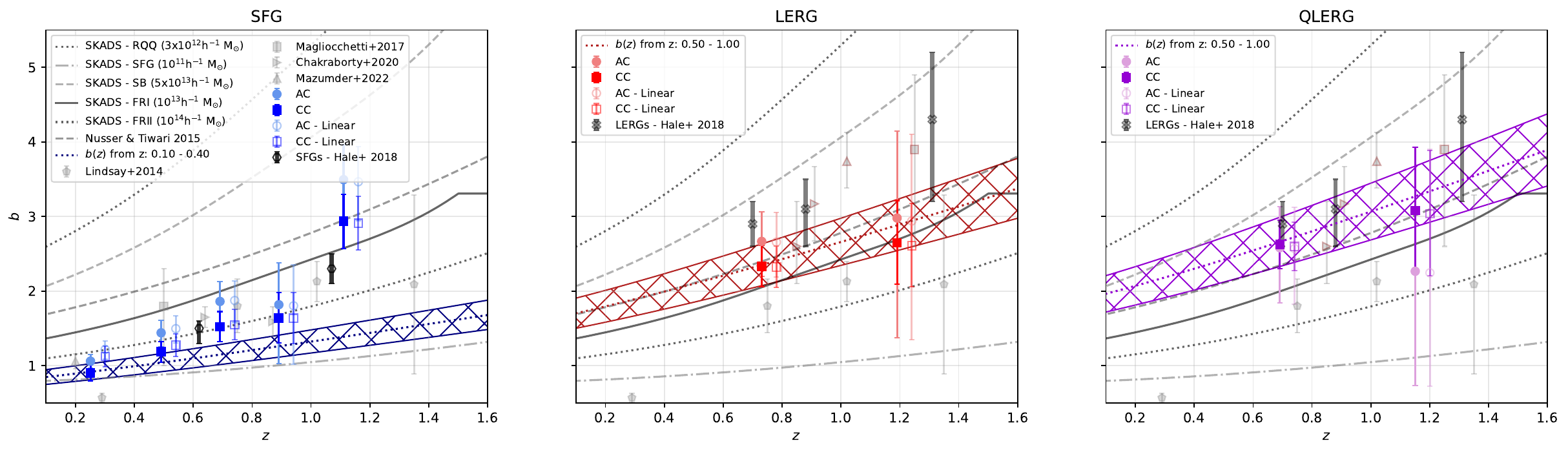}
        \caption{{Comparisons of $b(z)$ for SFGs (left), LERGs (centre) and QLERGs (right). {Filled light} colours indicate results from the auto-correlation function {(circles)} and dark colours {(for red, blue and purple)} indicate the results from the cross-correlation function {(squares)}, when using the `HaloFit' function. Additionally, artificially offset by $\delta z$=0.05 and semi-transparent are the results from using the `linear' model. We also show previous results of \protect \citet[dark grey dashed lines; this is for an AGN dominated population]{Nusser2015}, \protect \citet[grey pentagons; for sources not categorised by source type]{Lindsay2014}, \protect {\citet[grey squares]{Magliocchetti2017}},  \protect \citet[{grey right pointing triangles}]{Chakraborty2020}, \protect \citet[{grey upwards pointing triangles}]{Mazumder2022} and \protect \citet[black diamonds for SFGs and black crosses for LERGs]{Hale2018}. {For \protect \citet{Magliocchetti2017, Chakraborty2020} and \protect \citet{Mazumder2022} who measure the bias of AGN and SFGs separately, the points outlined in red represent the bias measurements for AGN and we only plot the source type relevant measurements in each given panel}. {The hatched regions indicate the evolutionary bias model that would be observed using the $b(z) \propto 1/D(z)$ model, using the bias in the lowest redshift bin for that source type. Additionally, the models used in \protect \cite{Wilman2008, Wilman2010} are also {shown as grey lines} for radio quiet quasars (RQQ, light grey dotted), star forming galaxies (SFG, light grey dot-dashed), starburst galaxies (SB, dark grey dashed), FRI galaxies \protect \cite[dark grey solid, see][for descriptions of FRI sources]{Fanaroff1974} and FRII galaxies (dark grey dotted).}}}
    \label{fig:bias_z}
\end{figure*}

\begin{figure}
    \centering
    \includegraphics[width=0.48\textwidth]{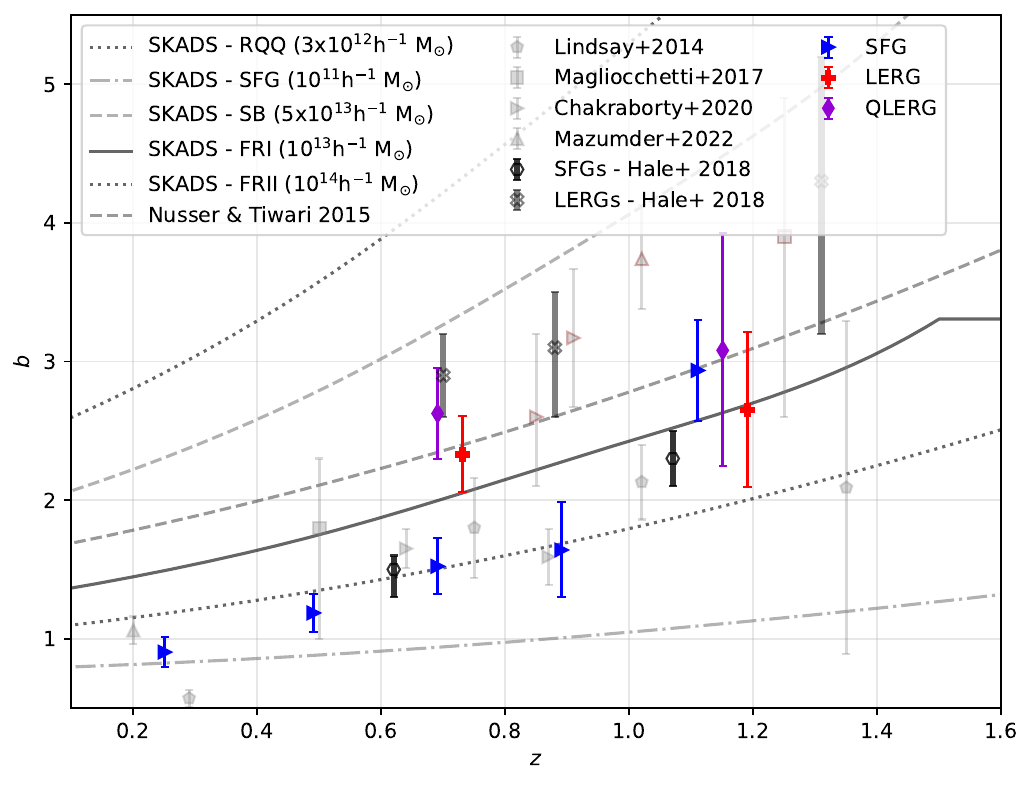}
        \caption{{As for Figure \protect \ref{fig:bias_z}, now showing the comparisons of bias between the different source {populations (SFGs, LERGs and QLERGs)}, using the bias derived from the cross-correlation function and using the `HaloFit' fitting model. {The previous models \protect \citep{Nusser2015} and data \protect \citep{Lindsay2014, Magliocchetti2017, Hale2018, Chakraborty2020, Mazumder2022} shown use the same plotting style as in Figure \protect\ref{fig:bias_z}.}} {Additionally, the models used in \protect \cite{Wilman2008, Wilman2010} are also {shown as grey lines} for radio quiet quasars (RQQ, light grey dotted), star forming galaxies (SFG, light grey dot-dashed), starburst galaxies (SB, dark grey dashed), FRI galaxies \protect \cite[dark grey solid, see][for descriptions of FRI sources]{Fanaroff1974} and FRII galaxies (dark grey dotted). The halo masses assumed for these populations are also shown in the legend.}}
    \label{fig:bias_z2}
\end{figure}

{For the star forming galaxies, our work} shows remarkable agreement across the redshift bins studied here to that of the studies of \cite{Hale2018}, \cite{Chakraborty2020} and \cite{Mazumder2022}. We measure a smooth evolution in the bias of SFGs, increasing from a bias, {{$b = \bccSFGOne^{+\bccperrSFGOne}_{-\bccnerrSFGOne}$ ($\bacSFGOne^{+\bacperrSFGOne}_{-\bacnerrSFGOne}$) for the cross-(auto-) correlation at the lowest redshift ($z\sim0.2$) to $b = \bccSFGFive^{+\bccperrSFGFive}_{-\bccnerrSFGFive}$ ($\bacSFGFive^{+\bacperrSFGFive}_{-\bacnerrSFGFive}$)}} at the highest redshifts considered ($z\sim1.2$). {An evolution in the bias for SFGs is in part expected, as there is evidence \citep[e.g.][]{Behroozi2013}, that halo masses of $\sim 10^{12}$ M$_{\odot}$ are the most efficient dark matter haloes for forming stars across a vast range of redshifts. As radio luminosity is known to be correlated to star formation rate \citep[see recent studies in e.g.][]{Davies2017, Gurkan2018, Smith2021}, we are likely observing highly efficient star forming galaxies. In order to reside in such a similar halo mass over cosmic time, this will require an evolution in the bias}. 

{{{In other LOFAR clustering studies that average} across all redshifts \citep[e.g.][]{Alonso2021,Hale2024, Nakoneczny2024}, we have assumed the bias is evolves inversely proportional to the growth factor, $b(z) = b_0/D(z)$.} {However, in this work we split into smaller redshift ranges than these previous studies and do not force $b_0$ to be the same in each redshift bin. Therefore, we are able to test whether this functional form is suitable to found in this work using smaller redshift bins. To do this we use the value of $b_0$ found} in the lowest redshift bin considered for the source population (SFG/LERG/QLERG) and trace its evolution under such a model. This is given by the hatched regions on Figure \ref{fig:bias_z}}. This comparison demonstrates that the bias values for SFGs are evolving at a more rapid rate than this previously assumed model, {with the evolutionary models used in \cite{Wilman2008}, suggesting that whilst the bias does increase with redshift \citep[as for the models of][]{Wilman2008}, the SFGs here are evolving at a quicker rate and with {larger bias} than for the `normal' SFG population of \cite{Wilman2008}. Assuming the models of \cite{Wilman2008}, our SFGs also suggest there may be some potential evolution above that for a constant halo mass. We note that \cite{Wilman2008} {split the SFGs into a starburst population, and a population of `normal' SFGs} galaxy population, whilst we do not distinguish the radio detected SFGs into sub-classes. However, our findings contribute to the growing evidence \citep[from e.g. studies of][]{Hale2018, Mazumder2022} that for a typical radio population at current sensitivities, using the bias models adopted in \cite{Wilman2008} to make predictions \citep[such as for cosmological predictions][]{Ferramacho2014, Raccanelli2012, Bacon2020} may not be appropriate (though see Section \ref{sec:limitations}). {Therefore, works such as \cite{Gomes2020} which adopt more recent bias measurement based models are key.} {Rapid evolution in bias has also been previously found for multi-wavelength studies of star-forming galaxies \citep{Magliocchetti2014} between $z\sim 1-2$.} 

This rapid evolution may relate to an intrinsic evolution for the star forming population, but may also relate to differences in the populations, where the higher redshift sources will typically be more luminous sources. {Therefore, a dependence of the bias on the radio luminosity of the source could drive an apparent evolution with redshift. This is investigated in Section \ref{sec:results_bLz}. } {However, such results may also be indicative of sources with increased AGN activity at higher redshift. Given the typically larger bias of our LERG population compared to the SFGs \citep[and more generally for AGN in the works of e.g.][]{Hale2018, Mazumder2022}, greater AGN contamination at the highest redshifts could increase the observed bias of SFGs. Indeed, there are different approaches taken to classify radio galaxies both on their multi-wavelength information such as through ultra-high resolution imaging \citep[see e.g.][]{Morabito2024} as well as using different SED fitting codes \citep[see e.g.][]{Das2024}. Both these approaches have been used to classify LOFAR data in these deep fields and {while the majority of classifications agree between the different methods, some differences are found in their in their proposed classifications to that of \cite{Best2023} which is used in this work. This includes at the higher redshifts for SFGs considered in this work.} This could be an alternative explanation for the agreement between the SFGs and LERGs in the highest redshift bins considered. However we do note that the bias measured for SFGs in the highest redshift bin is consistent with the evolution seen when the results of \cite{Hale2018} is combined in this work. }

For the LERG population, we measure a lower bias compared to the LERG analogues of \cite{Hale2018}. However, as noted in Section{ \ref{sec:data_class}, the QLERG population is believed to be a more direct comparison to the LERG population used in \cite{Hale2018}. In the current work, QLERGs show better agreement with the lower redshift work of \cite{Hale2018}, {though the LERGs are consistent within $\sim0.5\sigma$ to the QLERGs of this study and $\sim1\sigma$ to the work of \cite{Hale2018}}. Therefore we have weak evidence to suggest that quiescent LERGs reside in more biased haloes than the {general LERG population}. {This could suggest that similar to the wider population of galaxies, those with more significant star formation in the host galaxy appear to reside in less massive haloes. This would imply} that the underlying dark matter halo of a radio source may be influential in the properties of the radio source itself, or it appears at least related. The {bias of the} QLERG and LERG populations are more {uncertain though and so} could also be consistent with little-to-no evolution. 

Previous works studying the bias evolution in the LOFAR surveys {have typically been limited to higher flux density limits than considered in this work \citep[see e.g.][]{Alonso2021, Hale2024, Nakoneczny2024, Petter2024} by approximately a factor of $\sim$10 {and}} will be more dominated by AGN populations \citep[see][]{Best2023}. As such, for those brighter populations the $b(z)\propto 1/D(z)$ may have been an appropriate model for the bias. We note that in \cite{Hale2018}, it was noted that for the full AGN {population the bias appeared to flatten at the highest redshifts considered ($z\sim1.2-1.8$)}, which could be indicative of the downsizing of halos required to host equivalent sources at higher redshifts.} Supporting the results of \cite{Hale2018}, we also conclude that the bias models of \cite{Wilman2008} {for SFGs more closely reflect that assumed for their radio quiet quasar (RQQ) population {in this sample. \cite{Wilman2008} split the SFG populations into normal and starburst galaxies, therefore if such bias models are adopted for cosmological analysis \citep[e.g.][]{Raccanelli2012,Ferramacho2014} then a bias more representative for a realistic radio SFG population should be adopted. The halo mass estimates from \cite{Wilman2008} suggest such differences in the halo masses assumed for SFGs could be an order of magnitude and should be accounted for in order to place constraints on non-Gaussianity \citep[as updated in][]{Gomes2020}.} Studies similar to this work using deeper observations from precursor and pathfinder telescopes prior to the Square Kilometre Array Observatory (SKAO) are crucial to help understand the bias models to adopt in such studies}. Our results for LERGs suggest that {AGN sources, of this type} are more biased than {star formation dominated galaxies} up to intermediate redshifts ($z\lesssim 1$), but that the populations become more similar in their bias at higher redshifts. {This may be related to findings that the LERG populations appear to become dominated by star-forming hosts for $z\gtrsim 1$ in the luminosity functions of \cite{Kondapally2022}.}

\begin{table*}
    \centering
    \renewcommand{\arraystretch}{1.2}
    \begin{tabular}{c c c r r r r c c c c }
        Source Type & $z$ range & $z_{\textrm{med}}$ & N$_{\textrm{Radio}}$ & N$_{\textrm{Multi}}$ & $\log_{10}(M_{*} \ [M_{\odot}])$ & $N_{\textrm{R}}/N_{\textrm{D}}$ &$b_{{\textrm{AC}}}(z_{{\textrm{mid}}})$ & R-$\chi^2_{\textrm{min, AC}}$ & $b_{\textrm{CC}}(z_{\textrm{mid}})$ & R- $\chi^2_{\textrm{min, CC}}$\\ \hline \hline
SFG & 0.10 - 0.40 & \zmedSFGOne & \NDataRadioSFGOne & \NDataOpticalSFGOne & \MassMedSFGOne & \NrNdSFGOne & $\bacSFGOne_{-\bacnerrSFGOne}^{+\bacperrSFGOne}$ & \RedChiMinACSFGOne & $\bccSFGOne_{-\bccnerrSFGOne}^{+\bccperrSFGOne}$  & \RedChiMinCCSFGOne \\ 
SFG & 0.40 - 0.60 & \zmedSFGTwo & \NDataRadioSFGTwo & \NDataOpticalSFGTwo & \MassMedSFGTwo & \NrNdSFGTwo & $\bacSFGTwo_{-\bacnerrSFGTwo}^{+\bacperrSFGTwo}$ & \RedChiMinACSFGTwo & $\bccSFGTwo_{-\bccnerrSFGTwo}^{+\bccperrSFGTwo}$ & \RedChiMinCCSFGTwo \\ 
SFG & 0.60 - 0.80 & \zmedSFGThree & \NDataRadioSFGThree & \NDataOpticalSFGThree & \MassMedSFGThree & \NrNdSFGThree & $\bacSFGThree_{-\bacnerrSFGThree}^{+\bacperrSFGThree}$ & \RedChiMinACSFGThree & $\bccSFGThree_{-\bccnerrSFGThree}^{+\bccperrSFGThree}$ & \RedChiMinCCSFGThree \\ 
SFG & 0.80 - 1.00 & \zmedSFGFour & \NDataRadioSFGFour & \NDataOpticalSFGFour & \MassMedSFGFour & \NrNdSFGFour & $\bacSFGFour_{-\bacnerrSFGFour}^{+\bacperrSFGFour}$ & \RedChiMinACSFGFour & $\bccSFGFour_{-\bccnerrSFGFour}^{+\bccperrSFGFour}$ & \RedChiMinCCSFGFour \\ 
SFG & 1.00 - 1.30 & \zmedSFGFive & \NDataRadioSFGFive & \NDataOpticalSFGFive & \MassMedSFGFive & \NrNdSFGFive & $\bacSFGFive_{-\bacnerrSFGFive}^{+\bacperrSFGFive}$ & \RedChiMinACSFGFive & $\bccSFGFive_{-\bccnerrSFGFive}^{+\bccperrSFGFive}$ & \RedChiMinCCSFGFive \\ 
LERG & 0.50 - 1.00 & \zmedLERGOne & \NDataRadioLERGOne & \NDataOpticalLERGOne & \MassMedLERGOne & \NrNdLERGOne & $\bacLERGOne_{-\bacnerrLERGOne}^{+\bacperrLERGOne}$ & \RedChiMinACLERGOne & $\bccLERGOne_{-\bccnerrLERGOne}^{+\bccperrLERGOne}$ & \RedChiMinCCLERGOne \\ 
LERG & 1.00 - 1.50 & \zmedLERGTwo & \NDataRadioLERGTwo & \NDataOpticalLERGTwo & \MassMedLERGTwo & \NrNdLERGTwo & $\bacLERGTwo_{-\bacnerrLERGTwo}^{+\bacperrLERGTwo}$ & \RedChiMinACLERGTwo & $\bccLERGTwo_{-\bccnerrLERGTwo}^{+\bccperrLERGTwo}$ & \RedChiMinCCLERGTwo \\ 
QLERG & 0.50 - 1.00 & \zmedQLERGOne & \NDataRadioQLERGOne & \NDataOpticalQLERGOne & \MassMedQLERGOne & \NrNdQLERGOne & $\bacQLERGOne_{-\bacnerrQLERGOne}^{+\bacperrQLERGOne}$ & \RedChiMinACQLERGOne & $\bccQLERGOne_{-\bccnerrQLERGOne}^{+\bccperrQLERGOne}$ & \RedChiMinCCQLERGOne \\ 
QLERG & 1.00 - 1.50 & \zmedQLERGTwo & \NDataRadioQLERGTwo & \NDataOpticalQLERGTwo & \MassMedQLERGTwo & \NrNdQLERGTwo & $\bacQLERGTwo_{-\bacnerrQLERGTwo}^{+\bacperrQLERGTwo}$ & \RedChiMinACQLERGTwo & $\bccQLERGTwo_{-\bccnerrQLERGTwo}^{+\bccperrQLERGTwo}$ & \RedChiMinCCQLERGTwo \\ 

\end{tabular}
    \caption{{Summary table for fitting the bias from the auto- and cross-correlation for SFGs, LERGs and QLERGs within the redshift bins considered in this work. Included is the number of radio (N$_{\textrm{Radio}}$) and {multi-wavelength} sources (N$_{\textrm{Multi}}$) within the redshift bin, the median radio redshift in the bin (z$_{\textrm{med}}$), the median mass of radio sources in the bin from the consensus mass of \protect \cite{Best2023} and the ratio of randoms to data in the sample ($N_{\textrm{R}}/N_{\textrm{D}}$). Finally the bias from the auto-correlation ($b_{\textrm{AC}}$) and cross correlation ($b_{\textrm{CC}}$) {at the average redshift of the bin} alongside the minimum reduced-$\chi^2$ when fitting for $b$ is included. The bias results assume the Halofit model is used. }}
    \label{tab:bias_tab}
\end{table*}

\subsection{Luminosity dependence of bias for SFGs}
\label{sec:results_bLz}
{As discussed in Section \ref{sec:results_bz}, the bias of the SFGs appears to grow at a much faster rate than for the evolving model assumed in previous LOFAR studies \citep[][where $b(z) \propto 1/D(z)$ is assumed]{Alonso2021, Hale2024, Nakoneczny2024}. In this section we consider if this is driven by more luminous populations at higher redshifts, which are intrinsically more biased. This reflects work especially at other wavelengths such as that of \cite{Zehavi2011, Cochrane2017} and \cite{Clontz2022}. For the work of \cite{Zehavi2011}, their study of the clustering length, $r_0$, of blue galaxies compared to red galaxies shows an increase in $r_0$ with luminosity, whilst \cite{Cochrane2017} used H$\alpha$ detected SFGs at $z\sim0.8$ and found these populations to be more biased when more H$\alpha$ luminous populations were considered. We note, though, that \cite{Cochrane2023} appeared to observe a flattening in bias at larger H$\alpha$ luminosities for $z\sim1.5$ sources. However, the clustering of radio detected SFGs {as a function of} luminosity over a wide range of redshifts has not been studied in detail and can be limited by the redshift regimes probed by high- and low-luminosity samples \citep[see e.g.][]{Hale2018}. This is because large samples of SFGs from deep radio imaging {are} required, in regions where redshifts are available, such as {from} the LoTSS Deep Fields. {As discussed in Section \ref{sec:results_bz}, at radio wavelengths, the SFR and radio luminosities are known to be {well} correlated for star forming galaxies \cite[see e.g.][]{Garn2009, Davies2017, Gurkan2018, Smith2021}. If the {bias} of radio SFGs is correlated {with} the radio luminosity, this could in part explain the {bias evolution} as an effect of tracing different populations and more luminous star forming galaxies at the highest redshifts}. }

The LoTSS Deep Fields dataset is sufficiently large to allow us to {investigate whether we are able to constrain} how the bias of SFGs varies with both redshift and radio luminosity {simultaneously}. To do this we take the same approach as in the previous sections (where a given redshift range is selected) but additionally split into luminosity bins for each of the redshift bins that is considered. Specifically we use three luminosity bins for each of the redshift ranges considered, defined by taking the luminosities for sources with \texttt{Z\_BEST} values within the redshift range being considered and take the 33rd and 67th percentiles of the luminosities. For the $p(z)$ resampled data sets these will not be {exactly} even percentiles as the sources being considered in each redshift bin (and their luminosity) will vary, though should be approximately evenly distributed between luminosity bins. We {apply the same luminosity} cuts on the randoms using a combination of the redshift and the ``measured" integrated flux density to obtain {their luminosities}. As for the SFG sample where no luminosity cuts are applied, we compare the flux density, redshift and luminosity distributions of the data compared to the randoms, for which there is broad agreement, especially when the $p(z)$ resampled data are considered. These distributions are shown in Figures \ref{fig:flux_sfgs_z0p1_0p4} - \ref{fig:flux_sfgs_z1p0_1p3}.

The bias as a function of luminosity is presented in Figure \ref{fig:bias_L} and the measured values {are} given in Table \ref{tab:bias_l_tab}. This is given for both the auto-correlation and cross-correlation derived values. Such bias measurements are plotted at the median luminosity for sources with a \texttt{Z\_BEST} value in the {redshift and luminosity (from \texttt{Z\_BEST}) within the appropriate bin}. The results show broadly good agreement between the auto- and cross- derived bias values, though due to the smaller sample sizes being considered, the errors are larger for the auto-correlation and so challenging to draw any conclusions from. Therefore, our conclusions need to be drawn from the cross-correlation derived values. {The cross-correlation results in Figure \ref{fig:bias_L} show that any dependence of the median bias on luminosity is weak. To quantify this, we fit a simple linear model and using \texttt{scipy}'s \citep{scipy} \texttt{curve\_fit} module. {We} find slopes in the linear fit which are consistent with no evolution within {$\sim$1$\sigma$}. Therefore, we} {cannot comprehensively determine whether the differences in the luminosity {are} driving the evolution for SFGs seen in Figure \ref{fig:bias_z} or if the redshift evolution of bias is the only factor at play.} Larger source populations will be crucial for such studies which will be provided through deep surveys such as the second data release of the LOFAR deep fields \citep{Shimwell2025} and the {MIGHTEE survey \citep{Hale2024b}. }

\begin{figure*}
    \centering
        \begin{minipage}[b]{0.5\linewidth}
\includegraphics[width=8cm]{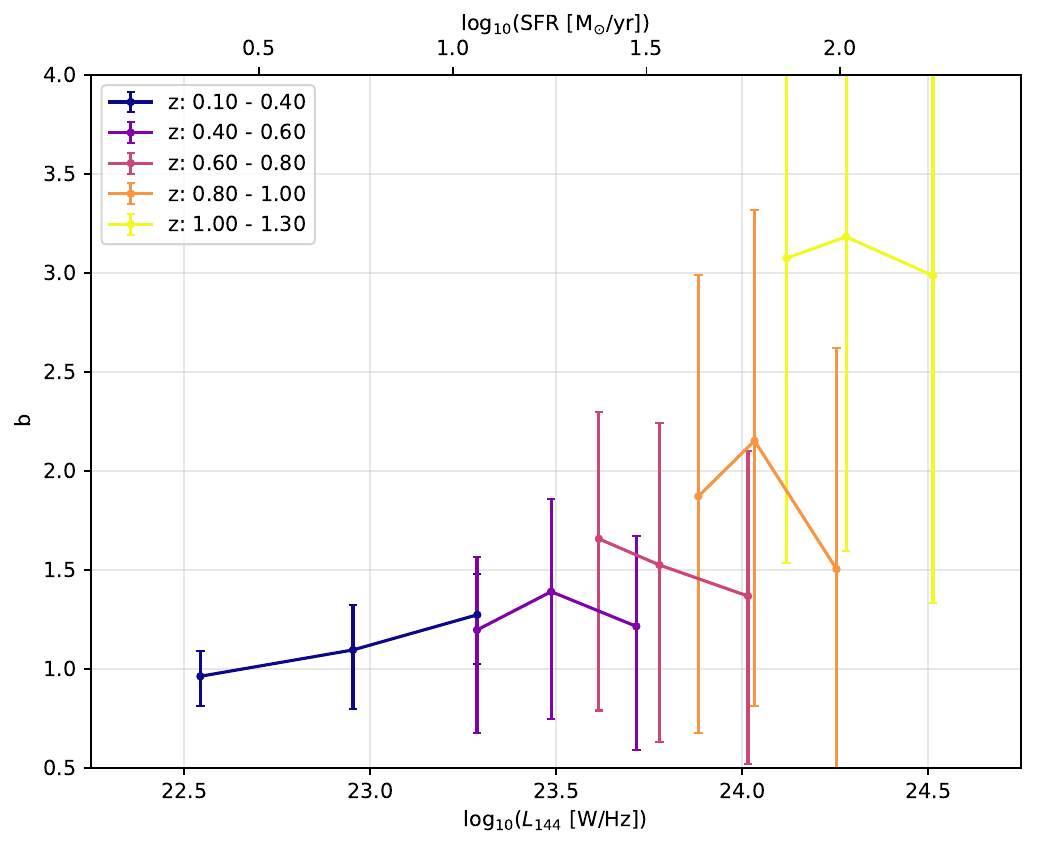}
     \subcaption{SFG - From Auto-correlation}
     \end{minipage}%
    \centering
        \begin{minipage}[b]{0.5\linewidth}
\includegraphics[width=8cm]{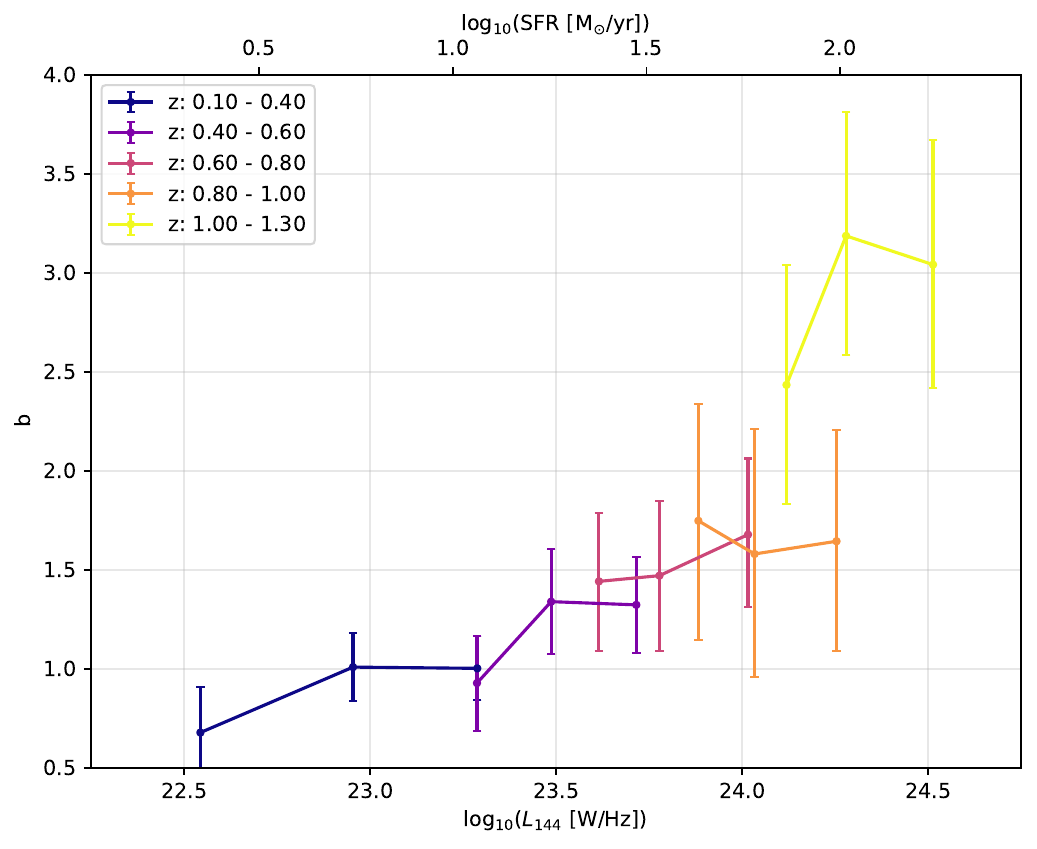}
     \subcaption{SFG - From Cross-correlation}
     \end{minipage}%
    \caption{{Bias as a function of luminosity for the auto- (left) and cross-correlations (right) of SFGs across the redshift bins considered in this work. Each colour represents a different redshift bin ranging from $z: 0.10-0.40$ (purple) to $z:1.00-1.30$ (yellow). The equivalent star formation rate (SFR) is also given on the top x-axis using the mass-independent conversion between luminosity and star formation rate of \protect \cite{Smith2021}.}}
    \label{fig:bias_L}
\end{figure*}

\begin{table*}
    \centering
        \renewcommand{\arraystretch}{1.2}
    \begin{tabular}{c c c c c c c c}
        $z$ & $\log_{10}$(L [W/Hz]) & Median & N & $b_{\textrm{AC}}(z_{\textrm{mid}})$  & R-$\chi^2_{\textrm{min}}$& $b_{\textrm{CC}}(z_{\textrm{mid}})$ &  R-$\chi^2_{\textrm{min}}$\\ 
        range & range & $\log_{10}$(L [W/Hz]) & & & & &  \\ \hline \hline
0.10 -0.40 & 21.74 - 22.79 & 22.54 & 2324 & $0.96^{+0.13}_{-0.15}$ & 0.05 & $0.68^{+0.23}_{-0.23}$ & 0.06 \\ 
0.10 -0.40 & 22.79 - 23.10 & 22.95 & 2307 & $1.09^{+0.23}_{-0.30}$ & 0.04 & $1.01^{+0.17}_{-0.17}$ & 0.06 \\ 
0.10 -0.40 & 23.10 - 24.79 & 23.29 & 2291 & $1.27^{+0.21}_{-0.25}$ & 0.22 & $1.00^{+0.16}_{-0.16}$ & 0.06 \\ 
0.40 -0.60 & 23.01 - 23.40 & 23.29 & 1705 & $1.20^{+0.37}_{-0.52}$ & 0.04 & $0.93^{+0.24}_{-0.24}$ & 0.21 \\ 
0.40 -0.60 & 23.40 - 23.58 & 23.49 & 1621 & $1.39^{+0.47}_{-0.64}$ & 0.05 & $1.34^{+0.26}_{-0.26}$ & 0.21 \\ 
0.40 -0.60 & 23.58 - 24.72 & 23.72 & 1717 & $1.21^{+0.46}_{-0.63}$ & 0.02 & $1.32^{+0.24}_{-0.24}$ & 0.21 \\ 
0.60 -0.80 & 23.42 - 23.70 & 23.61 & 1567 & $1.66^{+0.64}_{-0.87}$ & 0.06 & $1.44^{+0.35}_{-0.35}$ & 0.05 \\ 
0.60 -0.80 & 23.70 - 23.88 & 23.78 & 1562 & $1.52^{+0.72}_{-0.89}$ & 0.05 & $1.47^{+0.38}_{-0.38}$ & 0.05 \\ 
0.60 -0.80 & 23.88 - 25.16 & 24.02 & 1555 & $1.37^{+0.73}_{-0.85}$ & 0.06 & $1.68^{+0.38}_{-0.37}$ & 0.05 \\ 
0.80 -1.00 & 23.71 - 23.96 & 23.88 & 1174 & $1.87^{+1.12}_{-1.20}$ & 0.05 & $1.75^{+0.59}_{-0.60}$ & 0.15 \\ 
0.80 -1.00 & 23.96 - 24.12 & 24.03 & 1164 & $2.15^{+1.16}_{-1.34}$ & 0.03 & $1.58^{+0.63}_{-0.62}$ & 0.15 \\ 
0.80 -1.00 & 24.12 - 25.33 & 24.25 & 1165 & $1.50^{+1.11}_{-1.01}$ & 0.07 & $1.64^{+0.56}_{-0.56}$ & 0.15 \\ 
1.00 -1.30 & 23.93 - 24.20 & 24.12 & 1579 & $3.07^{+1.07}_{-1.54}$ & 0.03 & $2.43^{+0.60}_{-0.60}$ & 0.11 \\ 
1.00 -1.30 & 24.20 - 24.38 & 24.28 & 1581 & $3.18^{+1.12}_{-1.59}$ & 0.02 & $3.19^{+0.63}_{-0.60}$ & 0.11 \\ 
1.00 -1.30 & 24.38 - 25.79 & 24.51 & 1551 & $2.99^{+1.20}_{-1.65}$ & 0.03 & $3.04^{+0.63}_{-0.62}$ & 0.11 \\ 
 \hline
\end{tabular}
    \caption{{{Bias from the auto- ($b_{\textrm{AC}}$) and cross-correlation ($b_{\textrm{CC}}$) $\omega(\theta)$ for SFGs for different luminosity bins within the redshift bins considered in this work, evaluated at the mid point of the redshift bin. All luminosities are 144 MHz luminosities.} {We note that N is the number of sources in the luminosity bin based on \texttt{Z\_BEST} and is only indicative of the number of sources, as we use the resampled $z$ values from the $p(z)$ to make samples to make measurements of $\omega(\theta)$.}}}
    \label{tab:bias_l_tab}
\end{table*}

\subsection{{Limitations of this analysis}}
\label{sec:limitations}
{Whilst this work presented has placed constraint on the evolving bias of SFGs and LERGs within the LoTSS Deep Fields, there are limitations to the analysis, which we outline here {for completeness}. Firstly, {systematics may remain} that are unaccounted (or not fully accounted) {for} when obtaining the random catalogues of sources. {This} may impact the measurements of $\omega(\theta)$ and $b(z)$. However, there has been considerable effort to account for the observational systematics (see Section \ref{sec:randoms_radio}), so we believe remaining effects are less significant.} 

{Moreover, there are significant uncertainties of the redshift distributions for both the radio and multi-wavelength sources. {Uncertainty/variation in the $p(z)$} will affect measurements of bias (through the conversion of $\omega(\theta)$ to $b$). For example, {broader $p(z)$ models} was found to raise the bias, compared to if the \texttt{Z\_BEST} values were assumed to be correct. {This is because sources over a much larger redshift range require} a larger bias is needed to recreate the observed clustering {compared to} if they were accurately constrained within the redshift range in the bin. {Whilst we adopt} the redshift uncertainties of \cite{Duncan2021}, if these are over-estimated {the bias measurements} could be reduced. {Redshift uncertainty will be reduced with higher spectroscopic coverage for radio sources using surveys} such as WEAVE-LOFAR \citep{Smith2016}. }

{We also note that whilst {there are redshifts uncertainties for the data sources, the random catalogues} are idealised {and so do not have the same uncertainties in their redshifts.}. {However, what is important for the random sources is that they} reflect the observational detection across the fields. We {therefore }considered the impact of redshift uncertainties for the random catalogues by measuring $\omega(\theta)$ from using the random source catalogues in the neighbouring redshift bins (where available). This saw little change in the measured values of $\omega(\theta)$. Therefore, we believe our results are robust against the lack of uncertainties in the redshifts for the random catalogues.}

{Such redshift uncertainties {further affect the} multi-wavelength catalogues which we cross-correlate to. In our analysis, we choose to cut the optical galaxies based on their \texttt{Z\_BEST} redshift, with a mass limit applied (where the mass is determined assuming the best redshift). This allows for a consistent population for the radio galaxies to be correlated to. However, the large uncertainties in the redshift leads to a $p(z)$ with more dominance in broad wings, compared to {some} previous works \citep[such as][]{Hatfield2016, Shuntov2022}. This results in bias values for the optical sample with significant deviations to that of previous work. In order to test such effects, we considered the effect on the bias of the radio SFGs using the method adopted in this work, but also assuming the multi-wavelength redshift distribution is (i) the obtained from the redshift distribution of the \texttt{Z\_BEST} values and (ii) resampling the optical redshifts. In case (i), bias values for the multi wavelength catalogue were reduced, and {are} more comparable to \cite{Hatfield2016}, {and yet} we observe the same trend in the evolving bias for the radio selected SFGs. We note that in case (ii) it is computationally expensive to recalculate the stellar mass based on the new redshift and so we do not recalculate the mass of the sample. We again find the same trend in the bias evolution of the SFGs is recovered. For both cases the radio biases are within $\sim 1\sigma$ of the results presented in this work.  }

{Finally, this analysis will be improved in the future through full Halo Occupation Distribution (HOD) analysis \citep[as in e.g.][]{Zheng2005, Zheng2007, Hatfield2016}. The approach used in this work invokes a simpler approach of only fitting the large scale clustering with {a simple scaling for a} functional form of $\omega(\theta)$. Whilst this is different to the approaches of e.g. \cite{Hale2018, Chakraborty2020, Mazumder2022} who fit a power-law distribution, there are similarities in the approach that full HOD fitting is not used. Therefore, the approach in this work allows for a more similar comparison  to these previous works, without restricting ourselves to a power law model, which will not be appropriate across the range of angles considered. Our approach, though, may have differences to the effective bias found from HOD modelling {which accounts} for satellite galaxies within the samples to obtain halo mass estimates, halo properties and constrain bias values. Such relationships between halo mass and bias need to account for the full HOD in order to accurately probe halo masses, see discussion in \cite{Aird2021}. The combination of large radio samples and accurate redshifts such as WEAVE-LOFAR will, in future, allow more accurate constraints of the clustering evolution (and halo property evolution) for the dark matter environments hosting radio sources.  }

\section{Conclusions}
\label{sec:conclusions}
In this work we present a comparison of the clustering of star forming galaxies and low-excitation radio galaxies across the three LoTSS Deep Fields to trace both {their} evolution with redshift and the relationship between radio source populations and their underlying dark matter environments. We measure both the auto-correlation of the angular clustering of radio sources as well as the cross correlation with a catalogue of {multi-wavelength} sources across the fields, which total $\sim$26 sq. deg of combined area with deep multi-wavelength observations. By combining measurements of the angular two-point correlation function with knowledge of the redshift distribution within the fields assuming the full redshift distribution, $p(z)$, we obtain measurements of the galaxy bias (an indicator of how clustered galaxies are to dark matter) and traces its evolution to $z\lesssim1.5$ in a number of redshift bins. This evolution is measured both for sources separated as a function of source type, and for the SFG population also as a function of radio luminosity (a proxy for SFR).

Our work {suggests an} evolution in the bias for SFGs from {$b = \bccSFGOne^{+\bccperrSFGOne}_{-\bccnerrSFGOne}$ at $z$$\sim$0.2 to $b = \bccSFGFive^{+\bccperrSFGFive}_{-\bccnerrSFGFive}$ at $z$$\sim$1.2}. {This is at a quicker evolutionary rate than evolving bias model used for previous LOFAR studies of brighter populations \citep[with a more significant AGN population in e.g.][]{Alonso2021, Hale2024}, where $b(z) = b_0/D(z)$ and that this bias model may need to be modified for future work where broad redshift bins are considered.} {This may reflect a need for increasing mass halos to host star forming galaxies over cosmic time, however such rapid evolution at the highest redshift bin could also be indicative of either mis-classification of sources in the highest redshift bin (where AGN activity may actually be dominating the emission), or a luminosity dependence of the bias could be contributing to the rapid evolution seen in the bias at the highest redshifts studied, where in a flux limited surveys sources are naturally more luminous.} However the LERGs exhibit no such rapid evolution {({$b = \bccLERGOne^{+\bccperrLERGOne}_{-\bccnerrLERGOne}$ at $z$$\sim$0.7 to $b = \bccLERGTwo^{+\bccperrLERGTwo}_{-\bccnerrLERGTwo}$} at $z$$\sim$1.2)}, though are a factor of $\sim 1.5\times$ more biased compared to SFGs at lower redshift ($z\lesssim0.8$). This suggests that the dark matter haloes in which radio sources reside have a clear correlation to the radio populations {they host} {and that the haloes supporting SFGs may be less massive (by potentially an order of magnitude).}  We further consider the clustering of a subset of the LERG population known as quiescent LERGs (QLERGs), which do not have significant star formation contributions to their {overall emission}. These QLERGs have evidence that their bias may evolve {({$b = \bccQLERGOne^{+\bccperrQLERGOne}_{-\bccnerrQLERGOne}$ at $z$$\sim$0.7 to $b = \bccQLERGTwo^{+\bccperrQLERGTwo}_{-\bccnerrQLERGTwo}$} at $z$$\sim$1.2)}, {and} {weak evidence that they are} more clustered than the full LERG population at $z<1$. This bias evolution for LERGs and QLERGs is consistent the bias evolving inversely proportional to the growth function, however the uncertainties associated with such measurements means this {could also be weaker and consistent with potentially no,} evolution. 

Such differences {in the bias evolution of different source populations will likely} be important for future cosmology studies, such as with the SKAO, to exploit the differences in bias of the populations for cosmological studies \citep[e.g.][]{Ferramacho2014, Gomes2020}. {However, such studies need accurate models of the bias dependence of radio sources and so require studies with deep radio imaging where source classifications either through multi-wavelength source classifications \citep[e.g.][]{Whittam2022, Best2023} or through morphological classifications through high-resolution studies \citep[e.g.][]{Morabito2024}. Such studies would help disentangle the evolving bias evolution for different source populations and could also help understand more comprehensive dependencies of the radio populations on parameters intrinsic to the sources, such as their redshift, AGN activity, star formation rate and luminosity. To this end,} we consider the relationship of bias for SFGs on both the redshift and radio luminosity (a proxy for SFR) of the population being considered. This was in order to establish whether the rapidly evolving bias evolution for SFGs is as a direct result of observing typically more luminous populations when higher redshifts are considered. {We find that any luminosity-dependence of the bias is inconclusive, as whilst there is weak evidence at some redshifts for the best-fit bias to increase with luminosity, these results are not statistically significant. Therefore, it could instead be that the redshifts of the population are driving the evolution in bias.}. 

In the future, {spectroscopic surveys} such as WEAVE-LOFAR will help further address the question of the evolving relationships between radio sources and the underlying large-scale structure, allowing more accurate measurements of the redshift of sources and reducing the uncertainties introduced by the potentially {broad $p(z)$. Moreover {the combination of spectra alongside high resolution imaging will} help} to more comprehensively categorise sources and reduce potential classification errors. This combined with deeper radio data from the full LOFAR Deep Fields observations will improve our understanding of the galaxy-halo connection for radio sources.

\section*{Acknowledgements}

{We thank the referee for their helpful comments which have been important in improving the manuscript.} CLH acknowledges support from the Leverhulme Trust through an Early Career Research Fellowship. CLH and MJJ also acknowledges support from the Oxford Hintze Centre for Astrophysical Surveys which is funded through generous support from the Hintze Family Charitable Foundation.  PNB and RK are grateful for support from the UK STFC via {grants ST/V000594/1 and and ST/Y000951/1}. KJD acknowledges funding from the European Union's Horizon 2020 research and innovation programme under the Marie Sk\l{}odowska-Curie grant agreement No. 892117 (HIZRAD) and support from the STFC through an Ernest Rutherford Fellowship (grant number ST/W003120/1). MJJ acknowledges the support of a UKRI Fron- tiers Research Grant [EP/X026639/1], which was selected by the European Research Council, and the STFC consolidated grants [ST/S000488/1] and [ST/W000903/1]. DJBS acknowledges support from the UK STFC via grant numbers ST/V000624/1 and  ST/Y001028/1. We also thank D. Alonso for their advice with CCL.

LOFAR is the Low Frequency Array designed and constructed by ASTRON. It has observing, data processing, and data storage facilities in several countries, which are owned by various parties (each with their own funding sources), and which are collectively operated by the ILT foundation under a joint scientific policy. The ILT resources have benefited from the following recent major funding sources: CNRS-INSU, Observatoire de Paris and Université d'Orléans, France; BMBF, MIWF-NRW, MPG, Germany; Science Foundation Ireland (SFI), Department of Business, Enterprise and Innovation (DBEI), Ireland; NWO, The Netherlands; The Science and Technology Facilities Council, UK; Ministry of Science and Higher Education, Poland; The Istituto Nazionale di Astrofisica (INAF), Italy. The creation of LOFAR surveys data made use of the Dutch national e-infrastructure with support of the SURF Cooperative (e-infra 180169) and the LOFAR e-infra group. The Jülich LOFAR Long Term Archive and the German LOFAR network are both coordinated and operated by the Jülich Supercomputing Centre (JSC), and computing resources on the supercomputer JUWELS at JSC were provided by the Gauss Centre for Supercomputing e.V. (grant CHTB00) through the John von Neumann Institute for Computing (NIC). The creation of the LOFAR surveys data also made use of the University of Hertfordshire high-performance computing facility and the LOFAR-UK computing facility located at the University of Hertfordshire and supported by STFC [ST/P000096/1], and of the Italian LOFAR IT computing infrastructure supported and operated by INAF, and by the Physics Department of Turin university (under an agreement with Consorzio Interuniversitario per la Fisica Spaziale) at the C3S Supercomputing Centre, Italy.

This research made use of a number of tools and python packages: \texttt{Astropy}, community developed core Python package for astronomy \citep{astropy1, astropy2} hosted at \url{http://www.astropy.org/}; \texttt{TOPCAT} \citep{topcat1, topcat2}; \texttt{matplotlib} \citep{matplotlib}; \texttt{NumPy} \citep{numpy1,numpy2}; \texttt{SciPy} \citep{scipy}; \texttt{TreeCorr} \citep{Treecorr} and {\texttt{tqdm} \citep{tqdm}.}

\section*{Data Availability}
Data of the LoTSS Deep Fields is available through the LOFAR Surveys website \url{https://lofar-surveys.org} and further information of the data products can be found in \cite{Sabater2021, Tasse2021, Kondapally2021, Duncan2021} and \cite{Best2023}. Results presented in this work can obtained through a reasonable request to the author. 



\bibliographystyle{mnras}
\bibliography{clustering_deepfields} 




\appendix

\section{{Further validation plots for sources split by both radio luminosity and redshift}} 
{In Figures \ref{fig:flux_sfgs_z0p1_0p4} - \ref{fig:flux_sfgs_z1p0_1p3} we present the validation plots for SFGs split} into luminosity bins within a given redshift range. The rows are the same as used in Figure \ref{fig:flux_sfgs}, but now left to right indicates different luminosity ranges investigated.

\begin{figure*}
        \includegraphics[width=0.7\textwidth]{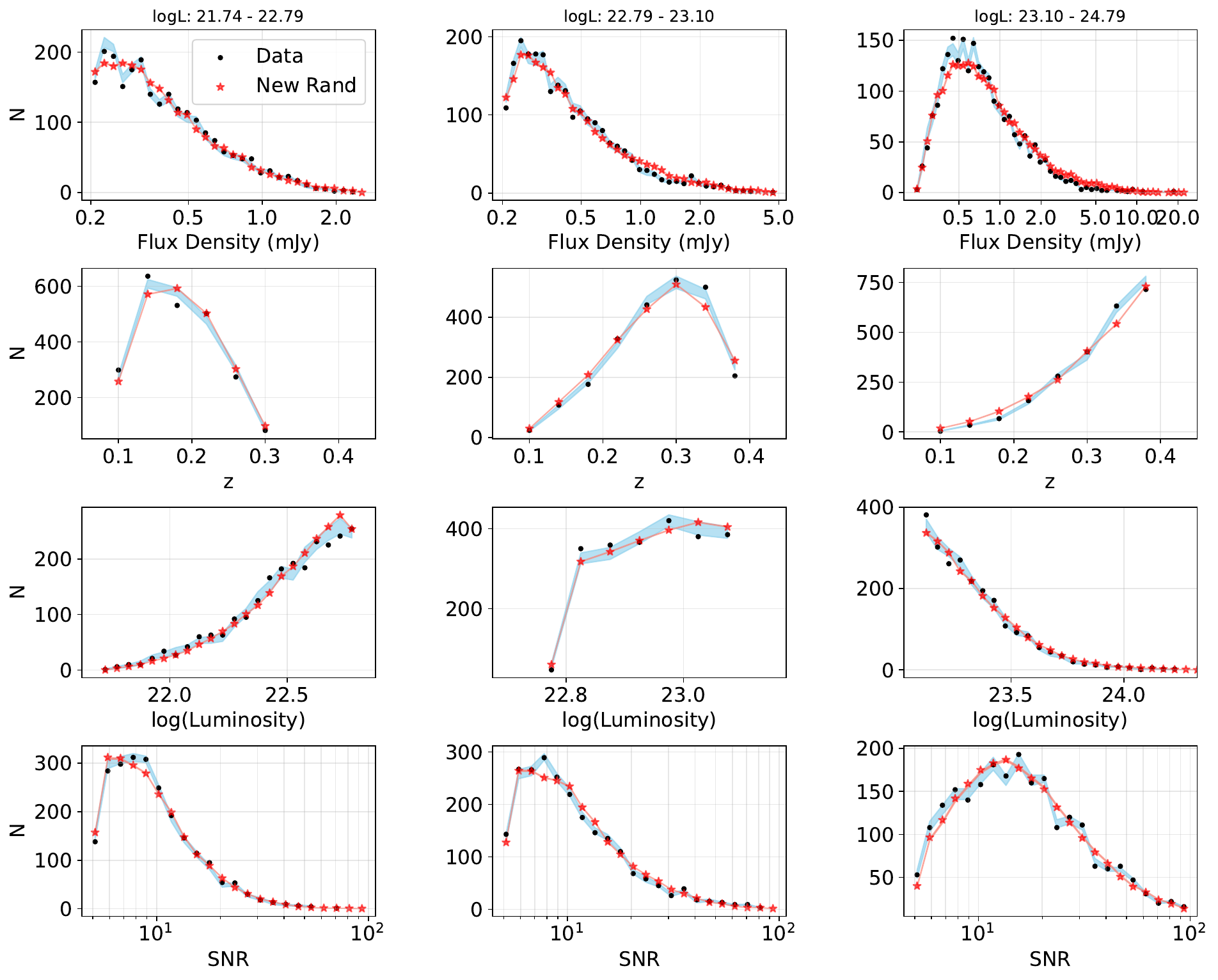}
    \caption{{As for Figure \protect \ref{fig:flux_sfgs} but for SFGs in the $z$: 0.1-0.4 redshift bin and then split into luminosity bins, increasing in luminosity from left to right.}}
    \label{fig:flux_sfgs_z0p1_0p4}
\end{figure*}

\begin{figure*}
        \includegraphics[width=0.7\textwidth]{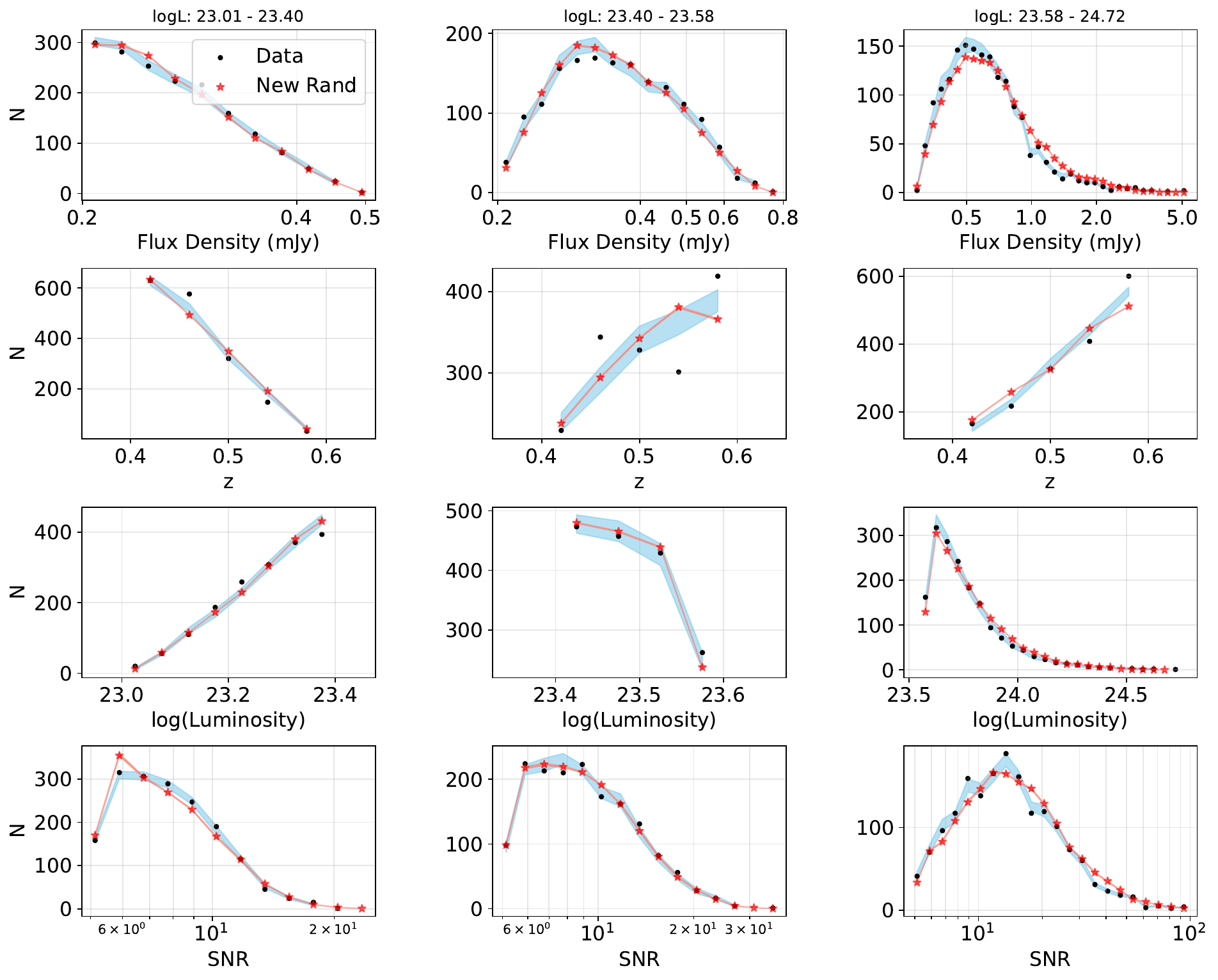}
    \caption{{As for Figure \protect \ref{fig:flux_sfgs} but for SFGs in the $z$: 0.4-0.6 redshift bin and then split into luminosity bins, increasing in luminosity from left to right.}}
    \label{fig:flux_sfgs_z0p4_0p6}
\end{figure*}

\begin{figure*}
        \includegraphics[width=0.7\textwidth]{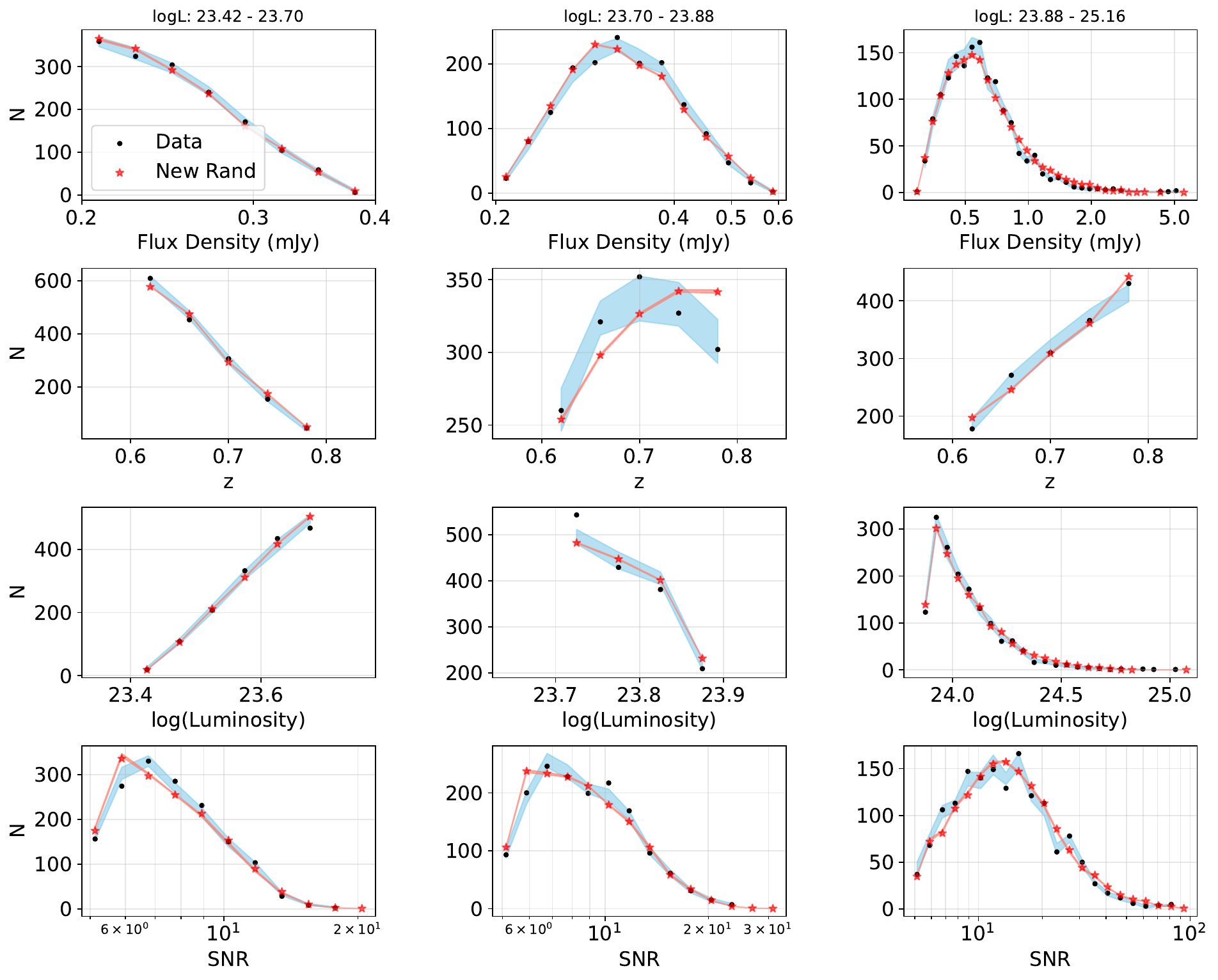}
    \caption{{As for Figure \protect \ref{fig:flux_sfgs} but for SFGs in the $z$: 0.6-0.8 redshift bin and then split into luminosity bins, increasing in luminosity from left to right.}}
    \label{fig:flux_sfgs_z0p6_0p8}
\end{figure*}

\begin{figure*}
        \includegraphics[width=0.7\textwidth]{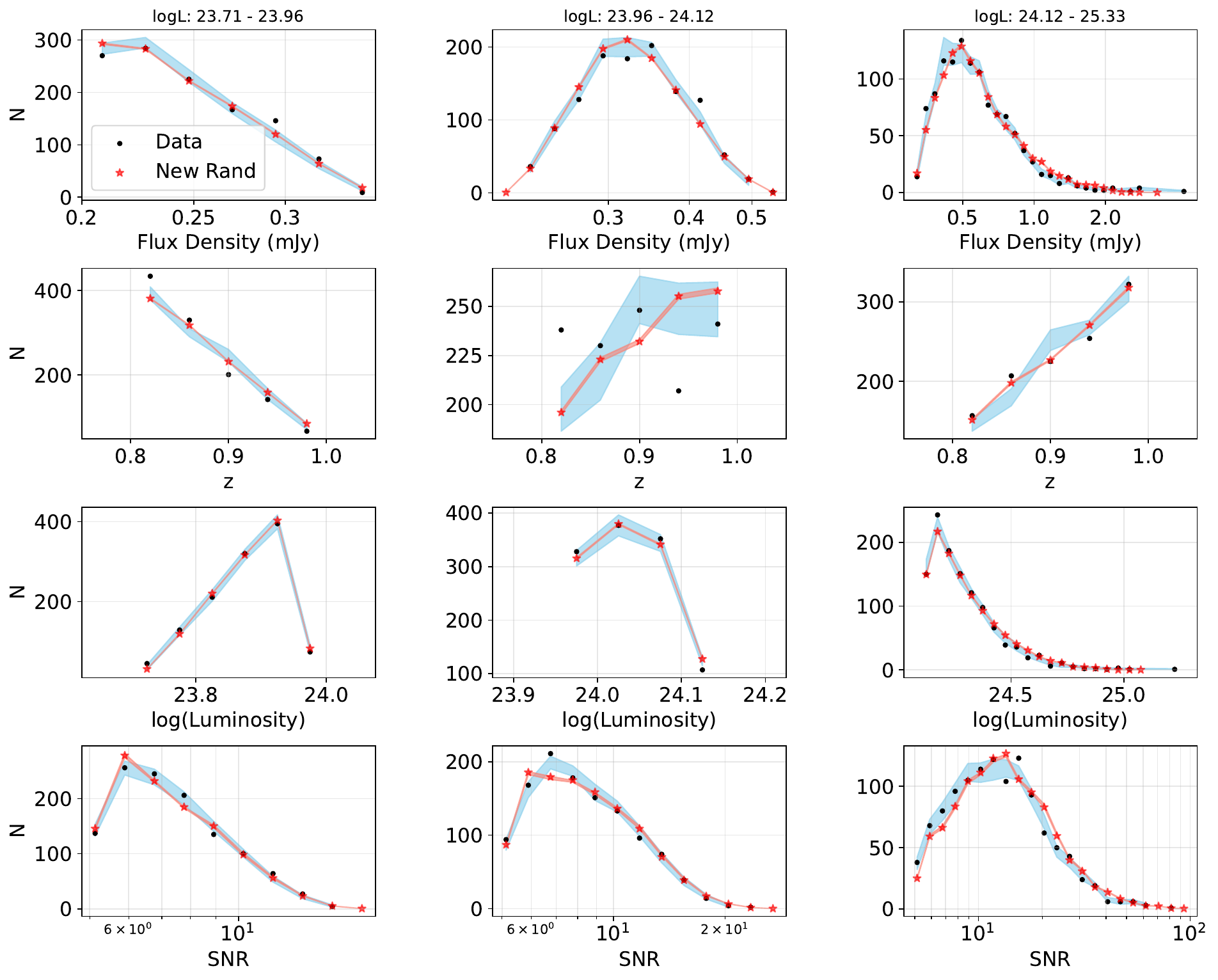}
    \caption{{As for Figure \protect \ref{fig:flux_sfgs} but for SFGs in the $z$: 0.8-1.0 redshift bin and then split into luminosity bins, increasing in luminosity from left to right.}}
    \label{fig:flux_sfgs_z0p8_1p0}
\end{figure*}

\begin{figure*}
        \includegraphics[width=0.7\textwidth]{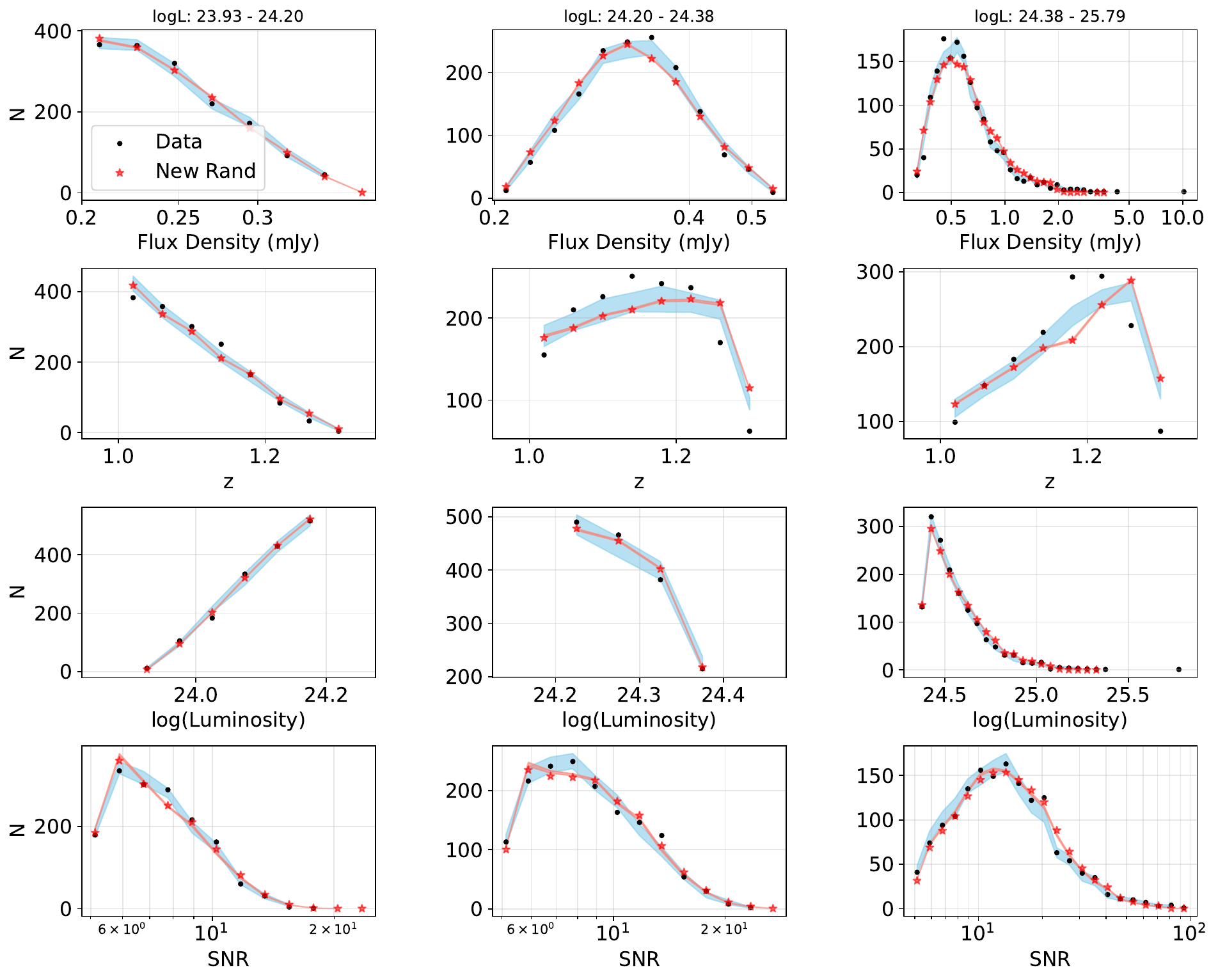}
    \caption{{As for Figure \protect \ref{fig:flux_sfgs} but for SFGs in the $z$: 1.0-1.3 redshift bin and then split into luminosity bins, increasing in luminosity from left to right.}}
    \label{fig:flux_sfgs_z1p0_1p3}
\end{figure*}

\bsp	
\label{lastpage}
\end{document}